# Colloidal Titanium Nitride Nanobars for Broadband Inexpensive Plasmonics and Photochemistry from Visible to Mid-IR Wavelengths

Sourav Rej,[a] Eva Yazmin Santiago,[b] Olga Baturina,[c] Yu Zhang,[a] Sven Burger,[d] Štěpán Kment,[a] Alexander O. Govorov*,[b] and Alberto Naldoni*,[a,e]

[a]Czech Advanced Technology and Research Institute, Regional Centre of Advanced Technologies and Materials Department, Palacký University Olomouc. Šlechtitelů 27, 78371 Olomouc, Czech Republic.

[b]Department of Physics and Astronomy, Nanoscale and Quantum Phenomena Institute, Ohio University Athens, Ohio 45701, United States.

[c]Chemistry Division, United States Naval Research Laboratory, Washington, D.C. 20375, United States.

[d]Zuse Institute Berlin, 14195 Berlin, Germany, JCMwave GmbH, 14050 Berlin, Germany.

[e]Department of Chemistry and NIS Centre, University of Turin, 10125 Turin, Italy.

Corresponding author contact:alberto.naldoni@unito.it; govorov@ohio.edu

**Abstract**

Developing colloidal plasmonic nanomaterials with high carrier density that show optical resonances and photochemical activity extending from the visible to the mid-infrared (MIR) ranges remains a challenging pursuit. Here, we report the fabrication of titanium nitride (TiN) nanobars obtained using a two–step procedure based on a wet chemical route synthesis of $TiO_2$ nanowires and their subsequent high temperature annealing in ammonia flow. Electromagnetic simulations of the resulting TiN nanobars reveal a rich set of optical resonances featuring transverse, longitudinal and mixed transverse–longitudinal plasmonic modes that cover energies from the visible to MIR region. TiN nanobars decorated with Pt co-catalyst nanocrystals show enhanced photocatalytic hydrogen evolution activity in comparison to both isotropic TiN nanospheres of similar size and TiN nanocubes under near infrared excitation at 940 nm due to the enhanced hot electron generation. We also demonstrate that plasmonic TiN nanobars can be used for the detection of furfural molecular vibrations by providing a strong surface enhanced infrared absorption (SEIRA) effect in the MIR region.

**Introduction**

In modern nanotechnology, plasmonics provides probably the finest control over light-matter interactions using metallic nanostructures supporting in–phase oscillations of conducting electrons strongly coupled to and excited by incoming photons. A relevant property of plasmonic nanostructures is their ability to concentrate light (and electromagnetic near fields) in nanoscale volumes, enabling strong enhancement of many physical and chemical phenomena underlying non–linear optics,[1] sensing,[2] heat transfer,[3,4] and chemical reactions,[5–7] among others. Once the plasmonic excitation decays, the stored energy can be transferred to adjacent objects (solid materials, molecules, fluids) either through plasmonic non–thermal (hot) carriers or by the significant lattice heating reached after their thermalization.[5,8] A flurry of promising experiments leveraging plasmonic effects have demonstrated that plasmonics will have a transformative impact on the energy technologies of the future.[9] Notably, the excitation wavelength of localized surface plasmon resonances (LSPRs) in plasmonic nanostructures defines their operational energy range (and therefore their application) and depend on the carrier density of the employed material, on the shape/size of the nanoresonator, and on its dielectric environment. For instance, plasmonic photocatalysts working in the visible region of the electromagnetic spectrum usually employ Au, Ag, Al, and TiN. The latter shows gold–like optical properties in the visible region,[10] while presenting high chemical stability in harsh environments (i.e. strong acids), high hardness, and extreme resistance to temperature (being a refractory metal) that make TiN an outstanding candidate material for several energy applications.[11–13]

Apart from anisotropic Au nanocrystals like rods, plates and bipyramids,[14,15] LSPR absorption in the near infrared (NIR) region can also be achieved by introducing element vacancy in semiconductors such as copper chalcogenides ($Cu_{2-x}S$, $Cu_7S_4$),[16–18] tungsten oxides ($WO_{3-x}$, $W_{18}O_{49}$),[19–22] $MoO_{3-x}$,[23] and $Bi_2O_{3-x}$,[24] which have shown photocatalytic activity in the NIR region. In addition, colloidal degenerate indium nitride (InN) nanocrystals showed an LSPR absorption reaching the mid-IR (∼3000 nm) region.[25] The mid–IR wavelength range is relevant for health- and defense-related applications including trace-gas detection, heat-signature sensing, thermal auditing, mimicking and cloaking, and source and detector development.[26] More importantly for energy applications, plasmonic devices working the mid–IR range find use in chemical sensing (or to track reaction products), as the majority of molecular vibrations fall in this energy range and can be amplified by the plasmonic near fields through the surface enhanced IR absorption (SEIRA) effect.[2] However, suitable materials for the mid-IR range are usually obtained by designing metasurfaces using complex nanoantenna geometries and expensive fabrication methods.[27,28] It is indeed challenging to achieve colloidal plasmonic nanostructures with LSPR resonances in the mid–IR using metals with high carrier density.

To address this major limitation, here we report the synthesis of photochemically active plasmonic TiN nanobars showing a multitude of resonances extending from the visible to the mid-IR region. Full-wave electromagnetic simulations reveals in details the superior optical properties of these anisotropic nanostructures in comparison to isotropic nanoparticles of comparable size. We validate the use of the investigated plasmonic TiN colloids, by testing them in plasmon–enhanced hydrogen evolution under monochromatic NIR light irradiation at 940 nm and comparing the shape-dependent photocatalytic activity of Pt-loaded nanocubes, nanospheres and nanobars, demonstrating the action of a hot electron driven mechanism. Moreover, we show that plasmonic TiN nanobars can be employed to detect energy-relevant molecules such as furfural by providing a strong SEIRA effect.

**Results and Discussion**

The studied plasmonic TiN nanostructures were prepared starting from the hydrothermal synthesis of the precursor $TiO_2$ nanomaterials adopting modified reported methods.[29,30] In one case, we obtained first hydrogen titanate ($H_2Ti_3O_7$) nanowires, which were converted into $TiO_2$ nanowires upon calcination at 400 °C for 1 h (Fig. 1a and S1). In the second case, we prepared $TiO_2$ nanospheres composed by the assembly of $TiO_2$ nanosheets (Fig. S2). Next, TiN nanobars and nanospheres were achieved by conversion of the $TiO_2$ nanostructures via nitridation at 800 °C for 5 h (see *Methods* in Supporting Information). A complete structural and morphological characterization of $TiO_2$ and TiN nanostructures confirmed the complete conversion to titanium nitride in both cases (Fig. 1c and S3–S9). TiN nanobars show an average length, width and height of 1200, 220 and 40 nm, respectively (Fig. 1d and S5), featuring nanosized holes on their surface (Fig. 1e–k). In contrast, TiN nanospheres show an average diameter of 200 nm and consist of complex spherical superstructures featuring surface nanoridges (Fig. S3). TiN nanobars present higher specific surface area (55.9 $m^2\ g^{-1}$) than the TiN nanospheres (46.8 $m^2\ g^{-1}$) (Fig. S10). Formation of nanoholes on the surface of TiN nanostructures is due to the high temperature anion exchange and phase transition process, which can be explained recalling the Kirkendall mechanism in the presence of hot ammonia (Fig. S7). The dark field scanning TEM image and energy-dispersive X-ray spectroscopy (EDS) elemental mapping (Fig. 1f−k) of TiN nanobars homogenously decorated with Pt nanocrystals (functional for the photocatalytic experiments shown below, for further characterizations see also Fig. S11–S13) provided nanoscale details on the nanoholes present on their surface. Fig. 1i shows the presence of oxygen in the elemental mapping due to the presence of an ultrathin amorphous oxide layer composed of titanium oxynitride (TiON) and titanium oxide ($TiO_2$) on the surface of TiN nanobars, as evidenced by HRTEM (Fig. S14) and XPS (Fig. S15, S16 and Table S1). The native oxide layer is commonly formed on TiN nanostructures upon exposure to air and/or water during chemical processing.[31]

The UV–vis–NIR absorption spectra measured in dichloromethane show that for both the TiO$_2$ nanostructures the absorption is mainly limited to the UV region (Fig. 2a and S17), with a characteristic tail in the visible region due to the light scattering produced by the nanostructures in suspension. In contrast, both the TiN nanomaterials present the characteristic interband transitions feature up to ~500 nm,[31] after which a broad and intense plasmon peak rises reaching the maximum in the NIR region. Interestingly, after 1300 nm, the absorption spectrum of TiN nanospheres decline, whereas for the nanobars, it kept the same intensity even beyond 2200 nm. To retrieve absorption data in a wider energy range, we measured attenuated total reflection-Fourier transform infrared (ATR-FTIR) spectra of the plasmonic TiN nanostructures (Fig. S18), showing that the absorption extends to the entire mid-IR wavelength range. We think that the observed broadband character of absorption is due to the size dispersion of TiN nanostructures and to their partial aggregation in colloidal suspension. This conclusion was confirmed by us using theoretical calculations.

To assess precisely the plasmonic properties of the TiN nanobars, we performed electromagnetic field propagation using finite-element software (COMSOL and JCMsuite). Geometrical features were retrieved from TEM micrographs (for width and length, see Fig. S5) and AFM height profile (for thickness, Fig. 1d). The width (220 nm) and thickness (40 nm) of the nanobars were kept constant, while the length was the sweeping parameter (200, 400, 800, 1200 nm). To account for the random orientation of the plasmonic nanobars in solution, the calculated spectra were averaged over three orthogonal propagation directions of light and two polarizations, thus leading to spectra featuring transverse (T), longitudinal (L) and mixed transverse–longitudinal (T–L) plasmonic modes (Fig. 2b and Fig. S20–S23). When we considered even more incident light directions (12 to 60) the observation of an additional plasmonic peak at ~2500 nm was possible (Fig. S24). To account for the granular structure featured in the nanobars (Fig. 1), we implemented the simulations by using a homogeneous dielectric function derived within the effective medium theory, where the TiN and dichlomethane volume fractions were 0.7 and 0.3, respectively. More details on the calculations can be found in the Supporting Information. Notably, the adoption of an effective medium dielectric function resulted in predicting relevant differences in the optical properties of the TiN nanobars as compared to the case where a metallic dielectric constant was used (Fig. S25). In particular, for the case of hybrid TiN nanobars with 1200 nm length, we observed two main effects: (i) a red shift of the plasmonic peaks, and (ii) an increased absorption fraction over scattering, especially at mid–IR wavelengths. Fig. 2b shows the averaged mass extinction coefficient (MEC) of the TiN nanobars with different lengths. This parameter quantifies how easily light is extinguished by the volume (mass) of the material, and it was calculated by dividing the optical cross sections with the mass of the nanostructure (see Supporting Information).

Increasing the anisotropy of the nanostructure, i.e. length ≥ 400 nm, three clear kinds of plasmonic resonances appear in the spectra. The surface charge maps at peak wavelengths obtained for specific light propagation direction (and polarization) provides insights into the nature of those plasmonic resonance breeds, denoted here as L, T1, and T2 (Fig. 2c–h and S23). The intense peak in the visible (642 nm for length=1200 nm) is due to the mixed $T_2$–$T_1$ transverse plasmonic mode and appears at similar wavelengths for nanobars with different lengths. A similar behavior is followed by the strong extinction peak in the NIR region (1262 nm for length=1200 nm), which is due to the mixed $T_1$–L plasmonic mode. In contrast, the most intense peak is assigned to the purely L plasmonic resonance and its position can be tuned from 2154 to 5266 nm by varying the TiN nanobar length from 400 to 1200 nm, respectively. Interestingly, the plasmonic peaks show increased bandwidth at longer wavelengths, reaching 1000–3000 nm full widths at half maximum for the mid–IR resonance. The computed spectral structure of plasmonic resonances is amazingly rich, and each spectral feature on the computed spectrum can be assigned to a certain resonance of this TiN plasmonic "flute". In addition, we found that the absorption and scattering efficiencies of the TiN nanobars are, in general, comparable (Fig. 2i and 2j). We note that the absorption efficiency is the most relevant parameter for the photochemical conversion, since it directly describes the portion of the photon energy, which become absorbed, i.e., get converted into heat and hot electrons. Simulated optical properties and related discussion for TiN nanospheres in detail can be found in the Supporting Information (Fig. S26–S28).

Notably, these results highlight the possibility to design TiN plasmonic nanostructures with high anisotropy suitable for plasmonic applications in a wide range of operational energies starting from the UV and extending to the mid-IR (300 nm – 6 μm).

To demonstrate the potential of plasmonic TiN nanobars, we tested their application both in plasmonic photocatalysis in NIR region and in SEIRA in the mid–IR energy range. Firstly, we selected plasmon–enhanced $H_2$ evolution through dehydrogenation of ammonia borane as a model reaction to compare the photocatalytic activity of TiN nanobars with nanospheres and nanocubes using monochromatic NIR excitation at 940 nm. Detailed calculation about photothermal turnover frequency ($TOF_{photo}$), thermal turnover frequency ($TOF_{therm}$), and hot electron turnover frequency ($TOF_{hot-e}$) can be found in the Supporting Information. We recently showed that TiN nanocubes can efficiently drive this reaction under both solar irradiation and monochromatic visible light, upon functionalization with Pt nanocrystals. The generation of plasmonic hot electrons was mainly localized in the TiN nanostructure (followed by their injection into the Pt component), while a smaller fraction of hot carriers was directly generated in the Pt nanocrystals.[31] Consequently, we functionalized the prepared TiN nanostructures with homogeneously dispersed Pt nanocrystals with an average diameter of 2.6 nm (Fig. S11, S12 and S30). As a result, we obtained TiN/Pt nanohybrids where the TiN component dominates the plasmonic properties, whereby Pt is the catalytic centre. The pure TiN nanostructures did not produce any considerable $H_2$ amount both under dark and light conditions. In contrast, we detected rapid $H_2$ evolution when TiN/Pt nanohybrids were employed as the photocatalysts (Fig. S31 and S32). The enhanced activity observed under light irradiation is attributed to plasmonic effects. The Pt nanocrystals themselves did not show any significant contribution as a photocatalyst during the light-assisted $H_2$ evolution, as confirmed by the identical reaction kinetics under dark and light conditions obtained by using Pt nanocrystals supported on an optically inactive $Al_2O_3$ support (Fig. S33). To determine the advantage of using the anisotropic TiN/Pt nanobars resonators in plasmonic catalysis, we compared their activity with Pt loaded TiN nanospheres and nanocubes. Using light excitation at 940 nm at a power density of 24 mW cm$^{-2}$, we measured TOF$_{hot-e}$ for different TiN shapes (Fig. 3a) in the experimental regime under constant temperature, i.e., when the low intensity light irradiation did not produce any significant macroscopic variation (and microscopic, see below and Supporting Information)[31–33] of the solution temperature. In this case, we avoid the thermal chemical mechanism and deal only with hot electrons. Notably, using the same Pt loading, TiN nanobars showed 1.67 and 2.41 times higher plasmon–enhanced $H_2$ evolution activity than that showed by TiN nanospheres and nanocubes respectively (Fig. 3a, S31, S32 and Table S4). The performance of TiN nanobars can be further highlighted by calculating the apparent quantum yield (for details see Supporting Information), which was 120%, 71% and 49% for TiN nanobars, nanospheres, and nanocubes respectively. TiN/Pt nanobars show also excellent NIR photocatalytic activity if compared with reported noble metal photocatalysts like Ag/$W_{18}O_{49}$ nanowires (AQY$_{1250nm}$ = 4%, see Table S5). To the best of our knowledge, this study provides first shape dependent plasmonic activity evaluation for TiN nanocrystals beyond 900 nm. Such unique NIR activity of TiN nanobars makes it suitable candidate for maximum utilization of solar light as well as under NIR (700−2500 nm) light. The TiN/Pt nanobars maintained their catalytic performance and structure even after three consecutive photocatalytic cycles (Fig. S34 and S35).

To investigate the reasons behind this observation, we computed the mass extinction coefficient for the absorption channel ($MEC_{abs}$) at 940 nm for TiN nanobars and TiN nanospheres with different sizes (Fig. 3b). TiN nanobars show generally higher MECs than TiN nanospheres, especially when the shape become more anisotropic (length ≥ 400 nm). Moreover, considering the optical cross sections at 940 nm (and the absorption/scattering efficiency reported in Fig. S29), TiN nanobars present similar values of both absorption and scattering cross sections, while for TiN nanospheres the scattering cross section is the double (or more for bigger sized nanospheres) the absorption one (Fig. S25 and S26). Hence, hot electron generation in nanobars can take place more efficiently due to the higher absorption fraction (i.e. less scattering), thereby resulting in higher $TOF_{hot-e}$ value, which was calculated as $TOF_{hot-e}= TOF_{photo} -TOF_{thermo}$.

The efficiency of TiN/Pt nanobars were further examined under higher light intensities (producing appreciable heating) and at corresponding different final reaction temperatures (FT) in the dark (Fig. S36, Table S4). An in–depth kinetic study revealed a steady increase of $TOF_{photo}$ with increasing light intensity (Fig. 3c). At 318 mW cm$^{-2}$ power density, $TOF_{photo}$ reached 242.3 $mol_{H2}\ mol_{Pt}^{-1}min^{-1}$ (see Table S5 for performance comparison with other photocatalysts reported in literature), with TiN/Pt nanobars showing a ∼3.1-fold rate enhancement with respect to dark room temperature conditions because of the combined effect of both photothermal heating and hot electrons. Furthermore, the Supporting Information includes a list of FTs for different light intensities and for various TiN nanosystems (Table S5) and estimates of local temperature increase at the surface of a single nanostructure (Section: Theoretical modelling). As expected, the local temperature increase at a nanoparticle is tiny, and the regime of photoheating in our study is collective, in which a large collective of nanocrystals contributes to the FT.[31–33]

Interestingly, the difference between $TOF_{therm}$ (Fig. 3c, blue bars) and $TOF_{photo}$ (Fig. 3c, red bars) increased with increasing light intensity, suggesting a prominent hot electron effect at higher light intensity.[7,31,34–36]

To confirm that a mechanism mediated by hot electrons is active even at high power density values, we computed Kinetic Isotope Effect (KIE) by comparing $H_2$ evolution rates obtained using $D_2O$ and $H_2O$ (Fig. S37 and 3d, see Supporting Information for more details on KIE calculation). The KIE value obtained in dark is 1.8, which confirms that the breaking of O−H bond in $H_2O$ is the rate–determining step (RDS) during the ammonia borane dehydrogenation. KIE obtained under light illumination is 2.4, which is higher as compared to dark conditions, confirming that photocatalytic $H_2$ evolution is a hot electron-driven process.[31,37]

The plasmonic effect on the activation energy ($E_a$) was also evaluated using the Arrhenius equation (Fig. 3e). Under dark, $E_a$ was 36.6 kJ mol$^{-1}$ (0.37 eV), while it decreased to 9.5 kJ mol$^{-1}$ (0.09 eV) under 940 nm light irradiation. This confirms that plasmonic excitation of TiN/Pt nanobars produce hot electrons, which directly participate in the RDS of $H_2$ evolution reaction (i.e., O−H bond cleavage) and stabilizes the transition state (TS) by 0.28 eV (Fig. 3f and Fig. S38).[31] Notably, this energy barrier reduction is higher than that one (0.1 eV, see Table S5) recently reported in a single-particle investigation study using Pt-tipped Au nanorods for the same chemical reaction. Moreover, we compare the energy barrier reduction for a selection of different reactions catalyzed by plasmonic photocatalysts recently appeared in the literature (Table S6). Despite some plasmonic photocatalysts provided impressive energy barrier reduction between 0.5 and 0.94 eV, most of them decreased the activation energy by 0.1−0.3 eV, similarly to the performance observed for TiN/Pt nanobars.

Next, the TiN nanobars were tested in the mid-IR range for SEIRA enhancement by measuring spectral features of furfural, a relevant biomass derivative employed in chemical industry for the production of solvents, polymers fabrication, adhesives, pharmaceuticals, and also a promising platform for photocatalytic $H_2$ production and biodiesel synthesis.[38] We compared the SEIRA enhancement obtained using TiN nanobars by using as the reference the IR spectrum collected using both an inert substrate as $Al_2O_3$ and a common plasmonic material made by Au nanoparticles supported on $Al_2O_3$ (Fig. 4 and Fig. S39). In the former (blue line), the signal intensity is very low in the whole investigated energy range showing a clear peak only for the most intense furfural vibration at 1670 $cm^{-1}$, which is assigned to the aldehyde group stretching. In the latter (black line), the signal is slightly increased over the whole wavenumber range with many peaks arising and being more defined due to SEIRA effect produced by the plasmonic Au nanoparticles. In contrast, the spectrum measured using the plasmonic TiN nanobars substrate (red line) almost all the spectral features that can be also found in the reference spectrum of pure furfural (black line). Notably, the SEIRA enhancements computed for all the detected peaks (Table S7) reach maximum values above 200 and span a vast energy range from 3000 to 13000 nm. These results are possible due to the broadband near fields generated in the plasmonic films made by aggregated TiN nanobars and that are shown to be superior that those generated in the reference sample containing Au nanoparticles.[2]

To elaborate more on the SEIRA effect, we expect that with the SEIRA enhancement may correlate with the scattering efficiency of a nanobar. We now see from Fig. 2j that the scattering contribution, which also describes the induced dipole moment on a nanobar, is nearly constant over a wide range of nanorod sizes. Therefore, the nanobars of various sizes can be efficient for SEIRA. The other interesting observation from Fig. S25 is that the scattering strength and the field-enhancement should depend on the composition of our nanobars. For the effective-medium model, the scattering becomes weaker (almost twice) since the hybrid material is less plasmonic.

**Conclusions**

In summary, we fabricated plasmonic TiN nanobars through high temperature nitridation of $TiO_2$ nanowires showing a rich spectrum of electromagnetic resonances extending from the visible range to mid-IR range. We demonstrated the superior photocatalytic activity of TiN nanobars for hot electron driven photocatalytic hydrogen evolution providing the first shape-dependent plasmonic activity evaluation for TiN nanocrystals using NIR excitation, while also showing the applicability and significant SEIRA enhancements of furfural molecular vibrations in the mid-IR. These results open the way to the utilization of anisotropic plasmonic TiN nanostructures as plasmonic substrates for enhancing chemical reactions and the detection of reaction intermediates/products for energy–relevant processes. Importantly, the developed colloidal nanostructures are made of an inexpensive, ceramic material, which is also highly thermally stable (refractory).

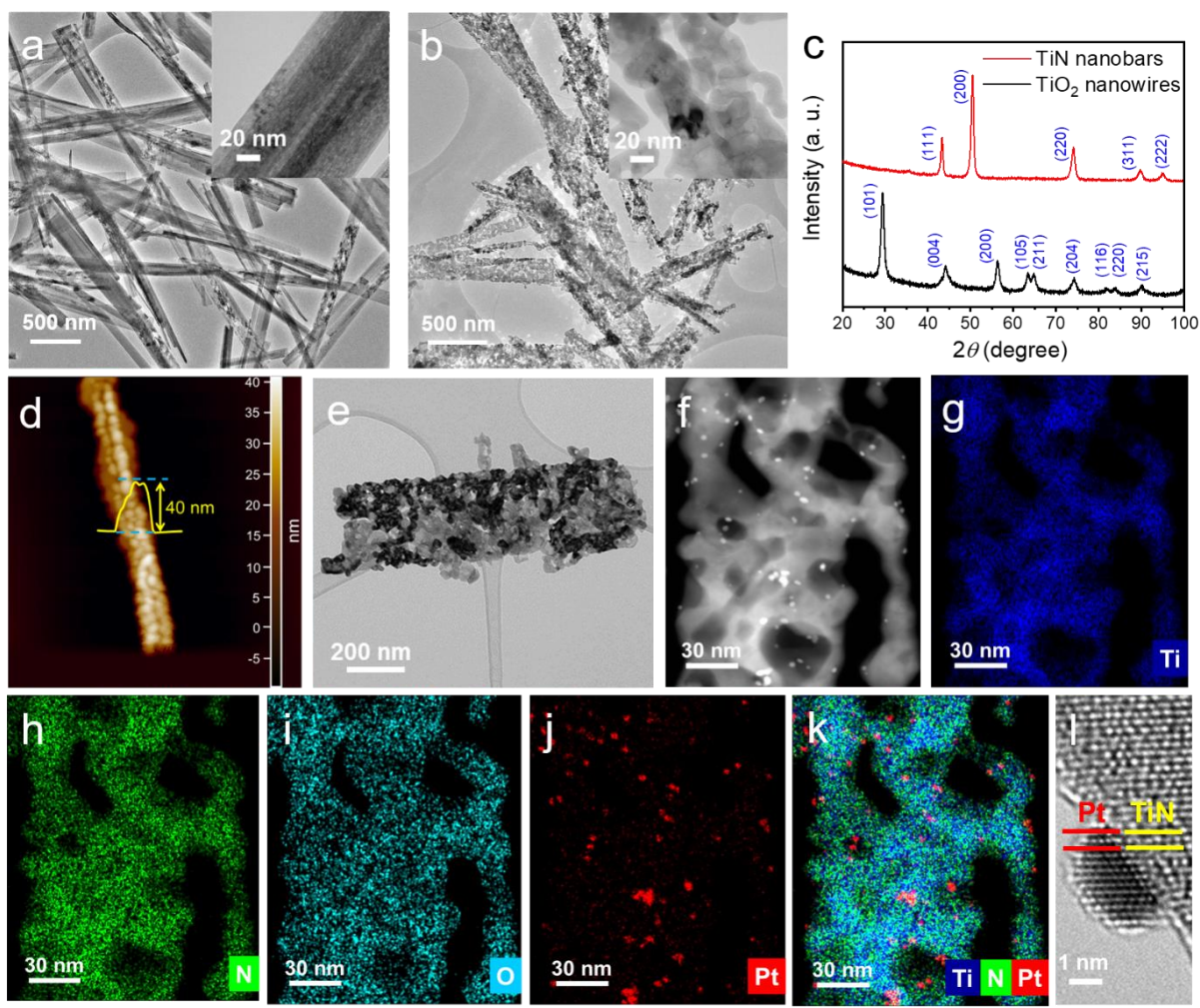

**Fig. 1** Large area and magnified TEM images (inset) of (a) TiO$_2$ nanowires and (b) TiN nanobars. (c) X-ray diffraction patterns of TiO$_2$ nanowires and TiN nanobars. (d) AFM height profile of a single TiN nanobar. (e) Representative TEM image of a single TiN nanobar loaded with Pt nanoparticles. (f) HAADF–STEM image and (g–k) EDS elemental maps of a single TiN/Pt nanobar. (l) HRTEM micrograph of TiN/Pt nanobars showing the TiN/Pt sharp interface.

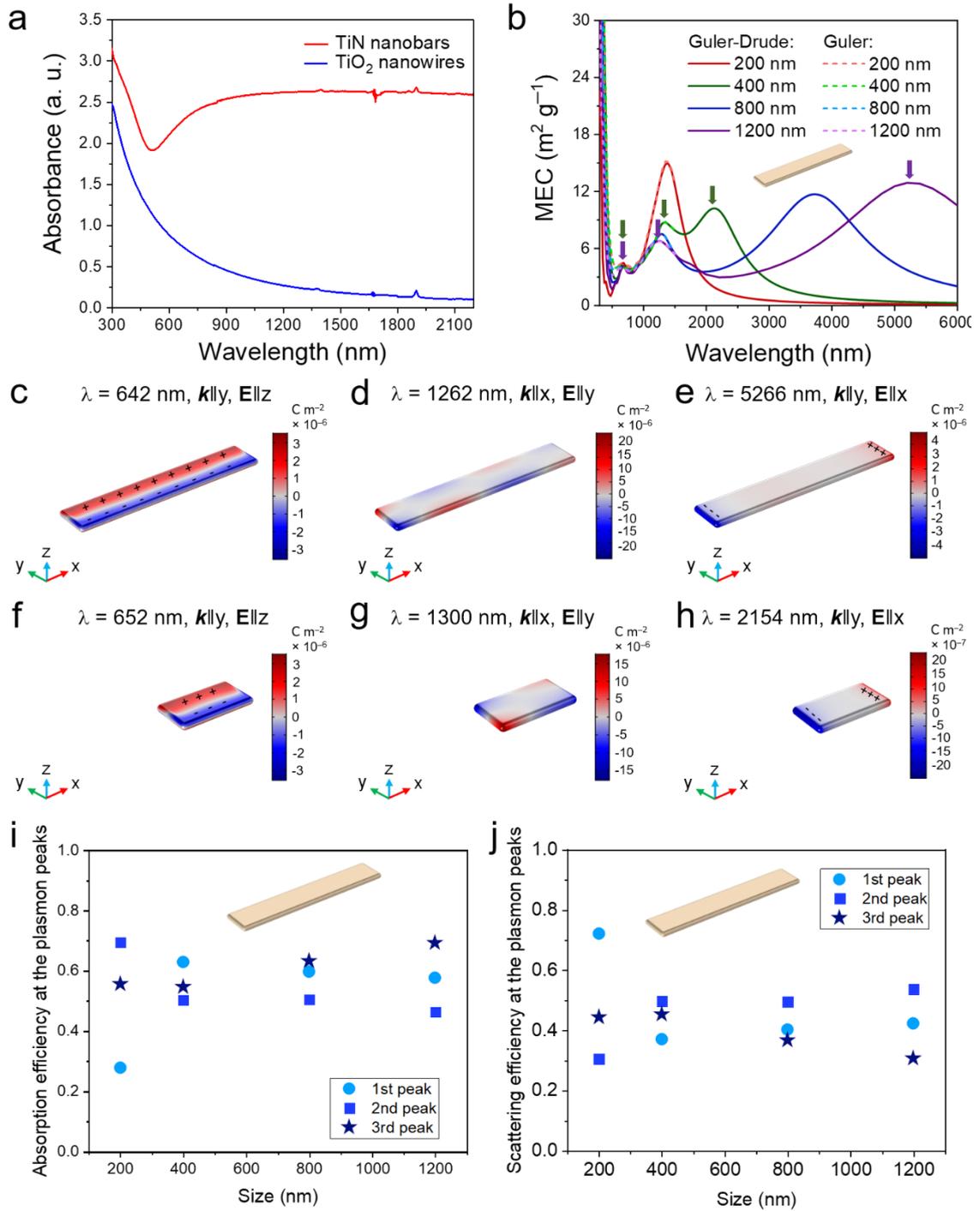

**Fig. 2** (a) UV-vis-NIR absorption spectra of TiO$_2$ nanowires (blue line) and TiN nanobars (red line) in dichloromethane. (b) Comparison of Mass Extinction Coefficient (MEC) for different sizes of nanobars structures. Effective medium theory. Non-normalized averaged 6-direction. W= 0.22 μM, H = 0.04 μM, $\varepsilon_m$ = TiN, $\varepsilon_i$ = 2.02892, $\delta_i$ = 0.3. Simulated surface charge maps of plasmons for nanobars with two different lengths (c–e) 1200 nm, (f–h) 400 nm, keeping width and height fixed. (i,j) Absorption and scattering efficiencies of TiN nanobars with different lengths.

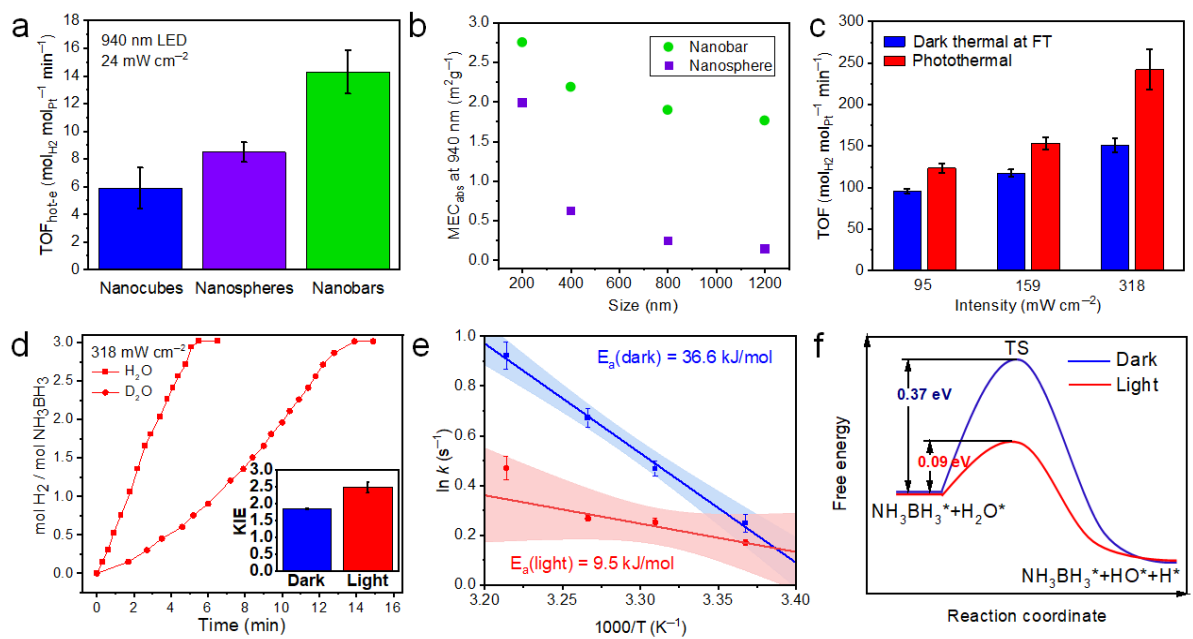

**Fig. 3** (a) Plasmon–enhanced $H_2$ evolution rates expressed as hot electron–driven turnover frequency (TOF$_{hot-e}$) for the TiN/Pt nanohybrids with different shape in the constant-temperature experimental regime. (b) Absorption mass extinction coefficient (MEC$_{abs}$) at 940 nm for TiN nanobars (green circles) and TiN nanospheres (purple cubes) with different sizes. (c) TOF of $H_2$ evolution over TiN/Pt nanobars in dark at final temperature (FT) and under 940 nm excitation at different power density. (d) Kinetics of $H_2$ evolution with TiN/Pt nanobars in $H_2O$ and $D_2O$. Inset shows Kinetic Isotope Effect (KIE) plot under dark and 940 nm light irradiation. (e) Arrhenius plots of apparent activation energies under light and dark conditions at FT showing coloured bands representing the 95% confidence interval of the linear fits. (f) Schematic representation of the free energy plots showing the energy barriers associated with the formation of the reaction transition state (TS) of the rate-determining step during $NH_3BH_3$ dehydrogenation under light and dark conditions. Asterisks denote species adsorbed on the catalyst surface. The error bar indicates the variation over three separate measurements.

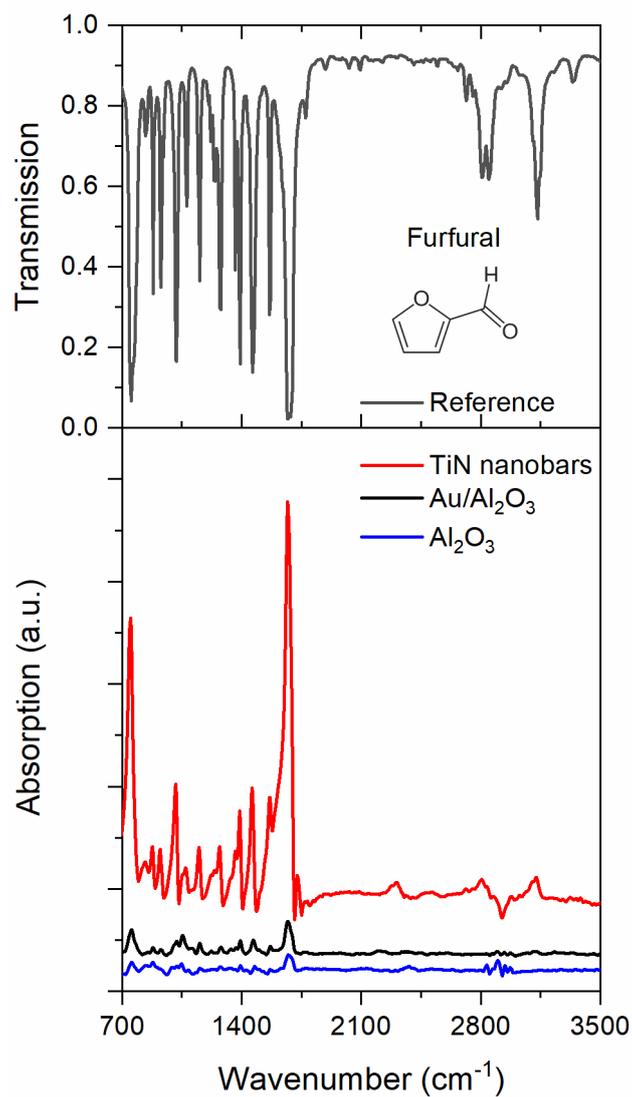

**Fig. 4** FT-IR reference spectrum of furfural taken from ref.[39] (grey) and SEIRA of furfural on Al$_2$O$_3$ (blue line), Au (6 wt.%)/Al$_2$O$_3$ (black line) and plasmonic TiN nanobars (red line).

**Acknowledgments**

SR, YZ, SK, and AN acknowledge the support of the Czech Science Foundation (GACR) through the award n. 20-17636S and the Ministry of Education, Youth and Sports of the Czech Republic and the Operational Programme Research, Development and Education - European Regional Development Fund, project no. CZ.02.1.01/0.0/0.0/15_003/0000416. EYS and AOG were funded via the ARO grant (contract W911NF-19-20081) and also supported by the Nanoscale and Quantum Phenomena Institute at Ohio University. AOG was also generously supported by the MATH+ Distinguished Visiting Scholarship Award, Berlin, and directly by Zuse Institute, Berlin. SB acknowledges funding by the Deutsche Forschungsgemeinschaft (DFG, German Research Foundation) under Germany´s Excellence Strategy (MATH+, EXC-2046/1, project ID: 390685689). OAB is grateful to US Naval Research Laboratory and Office of Naval Research for financial support.

# Supporting Information

**ethods**

*Synthesis of TiO$_2$ nanowires.* TiO$_2$ nanowires were synthesized using a modified protocol previously reported in the literature.[29] In a 100 mL plastic vial, 1 g of TiO$_2$ (anatase, 25 nm, Sigma-Aldrich) powder was dispersed in an aqueous NaOH solution (10 M, 30 mL) and sonicated for 5 min, followed by stirring (600 rpm) for 20 min and transferred into a 50 mL Teflon-lined stainless-steel autoclave. Once tightly sealed, the autoclave was placed into a preheated oven at 180 °C where it was kept for the next 72 h. Afterwards, the autoclave was taken out from the oven and cooled down naturally until reaching room temperature. The obtained white product (sodium titanate) was first washed with deionized water (DI-H$_2$O) for several times until the supernatant pH reached ~7 and then dried in a vacuum oven at 60 °C for 12 h. In the next step, this dried sodium titanate nanowire powder (1 g) was re-dispersed in hydrochloric acid (0.1 M, 100 mL) and stirred for 24 h at room temperature. The precipitate was separated, washed using the above mentioned procedure and dried under vacuum at 60 °C for 12 h. The obtained hydrogen titanate nanowires were then calcined in air at the heating rate of 2 °C min$^{-1}$. The obtained white product was labelled as TiO$_2$ nanowires.

*Synthesis of TiO$_2$ nanospheres.* TiO$_2$ nanospheres were synthesized using a modified protocol previously reported in the literature.[30] Using a 45 mL cylindrical vessel made by pressure-resistant glass, 2 mL TiCl$_3$ aqueous solution (~15%) were added dropwise into 30 mL ethylene glycol solution under stirring at 1000 rpm and let it stir for 5 min. After that, 2 mL of DI-H$_2$O were added dropwise under stirring and let it stir for another 5 min. Next, the vessel was tightly sealed and put into a preheated oil bath at 150 °C for 4 hours with constant stirring at 1000 rpm. Afterwards, the glass reactor was taken out from the oil bath and let it to cool down to room temperature naturally. The white product was then washed with DI-H$_2$O /ethanol mixture (1:1 volume ratio) and separated by centrifugation at 10000 rpm for 5 min for several times and finally dried in a vacuum oven at 60 °C for 12 h. These dried $TiO_2$ powder (200 mg) was re-dispersed in hydrochloric acid (0.1 M, 20 mL) and stirred for 24 h at room temperature. The precipitate was separated, washed using the above-mentioned procedure (but employing only DI-$H_2O$) and dried under vacuum at 60 °C for 12 h. The acid treated $TiO_2$ nanospheres were then calcined in air at 350 °C for 4 h using a heating rate of 1 °C min$^{-1}$. The as-obtained white product was labeled as $TiO_2$ nanospheres.

*Pseudomorphic conversion of $TiO_2$ into TiN nanostructures.* The as-synthesized $TiO_2$ nanowires and nanospheres (250 mg) were converted to TiN by nitridation at high temperature. Initially, Ar gas purged with a flux of 150 sccm for 15 min at room temperature to completely remove air from the quartz tube. Afterwards, the $TiO_2$ powders were heated at 2 °C min$^{-1}$ from room temperature to 800 °C under ammonia flow (150 sccm), and kept for 5 h at this temperature. The cooling process was carried out at 2 °C min$^{-1}$ under $NH_3$ flow to ensure complete conversion of $TiO_2$ to TiN.

*Synthesis of TiN/Pt nanohybrids.* The two different TiN nanostructures (70 mg) were decorated with Pt nanocrystals using a 25 mL round bottom flask and dispersing TiN in 4 mL of DI-$H_2O$ by 10 min sonication. The desired amount of $K_2PtCl_4$ was dissolved in 2 mL of DI-$H_2O$ added dropwise in the TiN solution under constant stirring at 600 rpm. After 3 h stirring, the mixture was transferred in a 10 mL plastic centrifuge tube and treated with a freeze drying process. The obtained dry powder was then reduced in a quartz tube furnace under a $H_2$/$N_2$ flow (10%:90% vol:vol, 200 sccm, heating rate 3 °C min$^{-1}$) for 2 h at 200 °C. The reduced TiN/Pt nanohybrids were kept in closed glass vials under an argon atmosphere before and after their use. Inductively coupled plasma mass spectrometry (ICP−MS) analysis confirmed that the Pt loading was 5 wt% for both TiN/Pt photocatalysts.

Using this procedure, commercially available 50 nm TiN nanocubes were also used to synthesize TiN/Pt hybrid nanocubes while keeping Pt loading fixed at 5 wt%.

In order to perform control experiments, a Pt/Al$_2$O$_3$ catalyst was also prepared using our previous reported procedure.[31] Pt nanocrystals were synthesized by reducing K$_2$PtCl$_4$ in the presence of NaOH (pH > 12) at 145 °C for 3 h in ethylene glycol. Then, this colloidal solution was vigorously mixed with Al$_2$O$_3$ followed by washing and overnight drying at 80 °C. Before using for catalysis, the as synthesized Pt/Al$_2$O$_3$ catalyst was reduced at 100 °C for 1 h under H$_2$ flow.

*Synthesis of Al$_2$O$_3$/Au nanocrystals.* Commercially available Al$_2$O$_3$ support was decorated with Au nanocrystals using ammonia borane reduction method under ice cold water. The desired amount of Al$_2$O$_3$ and HAuCl$_4$ was added to a 4 mL ice cold water and mixed via vigorous stirring for 15 min. Then 17 mg ammonia borane were dissolved in 1 mL DI-H$_2$O added quickly using a syringe through a rubber septum lid, followed by stirring for 15 min. Then as-synthesised Au/Al$_2$O$_3$ was cleaned 5 times with DI-H$_2$O (10000 rpm; 5 min), followed by overnight drying in an oven at 60 °C. These dried Al$_2$O$_3$/Au hybrid materials were then used for SEIRA sensing for furfural detection as compared to TiN nanobars. Au loading was 6 wt%.

*Characterization.* X-ray diffraction patterns of the material were determined using an X'Pert PRO MPD diffractometer (PANalytical) with iron-filtered Co Kα radiation (40 kV, 30 mA, λ = 0.1789 nm). X-ray photoelectron spectroscopy (XPS) analysis were performed on a PHI 5000 VersaProbe II XPS system (Physical Electronics) with a monochromatic Al Kα source (15 kV, 50 W) and a photon energy of 1486.7 eV. High-resolution spectra were scaled using the adventitious carbon peak as a reference. The low-resolution imaging of catalyst morphology was obtained with a transmission electron microscope (TEM) JEOL equipped with a LaB6 emission gun and operating at 160 kV. High-resolution micrographs and EDS elemental maps were acquired using a FEI Titan HRTEM microscope equipped with X-FEG electron gun operating at 80 kV. Visible-near IR (NIR) absorption spectra of the TiN, TiN-Pt and $TiO_2$ nanomaterials were measured using a Perkin-Elmer UV/VIS/NIR λ 1050 spectrometer. Prior to collecting spectra, TiN-based materials were dispersed in $CH_2Cl_2$ (Fisher Scientific, HPLC grade) by high-power ultrasound treatment using QSonica ultrasonic processor (model Q125) equipped with the sonicator probe (model CL-18). The samples were treated for 1 min (10 s ON, 10 s OFF) at 50% amplitude. Suspensions of $TiO_2$ nanospheres and nanobars in $CH_2Cl_2$ were homogenized for 5 min using a ultrasonic cleaner (Cole Parmer 8891). Spectra were collected in transmission mode. Spectra in the mid- and far-IR regions were acquired using the ATR module of the NikoletTM iS50 FTIR spectrometer. The powders were pressed against the all-reflective diamond window, and the FTIR spectra were collected.

*Hydrogen evolution experiments.* DI-$H_2O$ (22 mL) was added to 2.6 mg of TiN/Pt nanohybrids in a 25 mL round-bottom glass flask. The catalyst was dispersed by sonicating the solution for 1 min. The catalyst solution was then irradiated under 940 nm LED excitation for 5 min using a light beam size as large as the reactor cross section, with continuous stirring to homogenize the temperature in the solution. Different light intensities were calibrated with a thermopile detector (Standa 11UP19-H) before each experiment. Then, an aqueous ammonia borane ($NH_3BH_3$, Sigma-Aldrich) solution (0.27 mmol in 1 mL of water) was injected into the catalyst solution using a micropipette. The volume of evolved hydrogen gas was measured using an inverted water-filled burette connected to a sealed reaction flask through inserting a plastic tube. The volume of the evolved hydrogen gas was monitored by recording the displacement of water in the gas burette. Completion of the reaction was confirmed when gas generation stopped. The final reaction temperature (FT) was then measured with a thermocouple inserted into the reaction mixture. To evaluate the thermal contribution, the same reaction was carried out under dark conditions at FT.

*Surface enhanced infrared absorption (SEIRA) spectroscopy.* SEIRA spectroscopy measurements were conducted on an iS5 FTIR spectrometer (Thermo Nicolet) using the Smart Orbit ZnSe ATR (attenuated total reflection) accessory. All spectra were collected with nitrogen flowing the ATR accessory chamber. The spectrum of Air was collected as the background. The investigated SEIRA supports (TiN nanobars and $Al_2O_3$) were deposited over the ZnSe crystal surface and then pressed to form a compact layer. The spectra of the pristine supports were collected as a reference showing no IR bands. The SEIRA spectra of furfural were measured after dropping 2 μL of a 22 wt% furfural solution in ethanol onto the surface of the plasmonic TiN nanobars, Au (6 wt.%)/$Al_2O_3$, $Al_2O_3$, and waiting until the complete evaporation of the solvent. The furfural IR signals were extracted from the absorption spectra after baseline subtraction.

*Theoretical modelling.* The theoretical modeling of the TiN nanospheres and nanobars was mainly computed with COMSOL Multiphysics. The system consisted of randomly oriented porous nanoparticles in dichloromethane, which has a refractive index equal to 1.4244. The variation of size of the nanospheres is defined by the diameter and for the nanobars is the length, while the width and thickness are constant. These parameters are shown in Table S2.

To account for the random orientation of the particles in solution, the incident electromagnetic wave was computed from the average of three orthogonal propagation directions and two polarizations. Therefore, the absorption ($\sigma_{abs}$), scattering ($\sigma_{scat}$), and extinction ($\sigma_{ext}$) cross sections were then averaged in the following way:

$$\sigma_{\text{average}} = \frac{\sigma_{k\|x,E\|y} + \sigma_{k\|x,E\|z} + \sigma_{k\|y,E\|x} + \sigma_{k\|y,E\|z} + \sigma_{k\|z,E\|x} + \sigma_{k\|z,E\|y}}{6}$$

Additionally, by definition,

$$\sigma_{\text{ext}} = \sigma_{\text{abs}} + \sigma_{\text{scat}}$$

The metal dielectric function of TiN was based on the empirical dielectric function obtained by Guler et. al.[S1] This data does not, however, cover the entire frequency range relevant for this study. Therefore, a Drude model was used to fit and extrapolate the dielectric function to a larger set of frequencies. The Drude model is described by the following equation:

$$\varepsilon_{\text{metal}}(\omega) = \varepsilon_{\text{b,Drude}}(\omega) - \frac{\omega_p^2}{\omega(\omega + i\Gamma_D)}$$

where $\varepsilon_{\text{b,Drude}}$ is the Drude bulk dielectric function, $\omega_p$ is the Drude plasma frequency and $\Gamma_D$ is the Drude damping; their values were found to be 8.76, 6.94 eV and 0.297 eV, respectively, when fitting the empirical dielectric function to the equation. To account for the porosity of the nanoparticles, the Maxwell Garnet approximation was used as the effective medium theory. This homogenization theory was used to mix the dielectric function of TiN and dichloromethane, which fills up the nanoparticle pores, resulting in a homogeneous effective medium dielectric function. The Maxwell-Garnett equations is given by:[S2,S3]

$$\varepsilon_{\text{eff}} = \varepsilon_m \frac{2\delta_i (\varepsilon_i - \varepsilon_m) + \varepsilon_i + 2\varepsilon_m}{2\varepsilon_m + \varepsilon_i - \delta_i(\varepsilon_i - \varepsilon_m)}$$

where $\varepsilon_{\text{eff}}$ is the effective dielectric function of the medium, $\varepsilon_i$ is the dielectric function of the voids (filled by dichloromethane), and $\varepsilon_m$ of the metal; $\delta_i$ is the void volume fraction, which was taken as 0.3. All the dielectric functions previously described are shown in Fig. S19.

To further study the nanoparticles, the mass extinction coefficient (MEC) of the nanoparticle was calculated. This quantifies how easily light is extinguished by the volume of the material. Therefore, this quantity is a variant of the cross section, hence the light absorption and scattering is defined per mass rather than per particle. The MEC equations for extinction, absorption and scattering are expressed by:

$$MEC = \sigma_{\text{ext}}/mass$$

$$MEC_{\text{abs}} = \sigma_{\text{abs}}/mass$$

$$MEC_{\text{scat}} = \sigma_{\text{scat}}/mass$$

To calculate the mass, the density for TiN is $\rho_{TiN} = 5.4$ g cm$^{-3}$. Finally, surface charge maps were generated on COMSOL Multiphysics at specific wavelengths, defined by the plasmonic resonances.

The other optical parameters, used by us in the main text, are the absorption and scattering efficiencies:

$$\text{Eff}_{abs} = \sigma_{abs}/\sigma_{ext}$$

$$\text{Eff}_{scat} = \sigma_{scat}/\sigma_{ext}$$

The role of photoheating can be understood by computing the local phototemperature increase induced at the surface of single nanorod, $\Delta T_{local,surf} = T_{local,surf} - T_0$, where $T_{local,surf}$ and $T_0$ are the photoinduced and equilibrium temperatures, respectively. For the TiN nanobars of typical size of 1.2 μm x 0.22 μm x 0.04 μm, the estimate for the local temperature increase under the typical illumination (159 mW cm$^{-2}$, 940nm) is a really small number, $\Delta T_{local,surf} \sim 3*10^{-5} K$. The physical reasons for such a small number are in a very low level of illumination and a high thermal conductivity of the water matrix. For the case of the TiN nanosphere, the estimated excess temperature is also tiny, $\Delta T_{local,surf} \sim 10^{-4} K$ (D_sphere=200 nm). To obtain the above estimates, we used the following equation taken from Ref. 31:

$$\Delta T_{local,surf} \sim \frac{q_{tot,NC}}{4\pi k_{t,w}} \frac{1}{R_{eff}}$$

$$q_{tot,NC} = \sigma_{NC} \cdot I_0$$

Here $k_{t,w}$ is the thermal conductivity of water and $R_{eff}$ radius is a thermal effective radius of a nanocrystal (NC), which should be computed from the integral:

$$\frac{1}{R_{eff}} = \frac{1}{V_{NC}} \int_{V_{NC}} dV \frac{1}{\sqrt{x^2 + y^2 + z^2}}$$

Regarding the above estimates, we should note that the above equation for $\Delta T_{local,surf}$ is an approximate, however giving good agreement with the exact COMSOL numbers (Ref. 31). Our conclusion is that the photo-heating is our experiments is governed by the collective mechanism,[S4,S5] in which a large collection of NCs contributes to the resulting phototemperature. The local increase at the surface of a single NC is tiny, i.e., $\Delta T_{local,surf} \ll \Delta T_{FT}$. Indeed, the final FT at 159 mW cm$^{-2}$ for the TiN nanobars is $\Delta T_{FT} \sim 11K$, whereas the corresponding local temperature increase, as obtained from our estimate, is $\Delta T_{local,surf} \sim 3*10^{-5} K$. Finally, we note that this collective regime of plasmonic photoheating is very typical for solution experiments with NCs.

**Material Characterization**

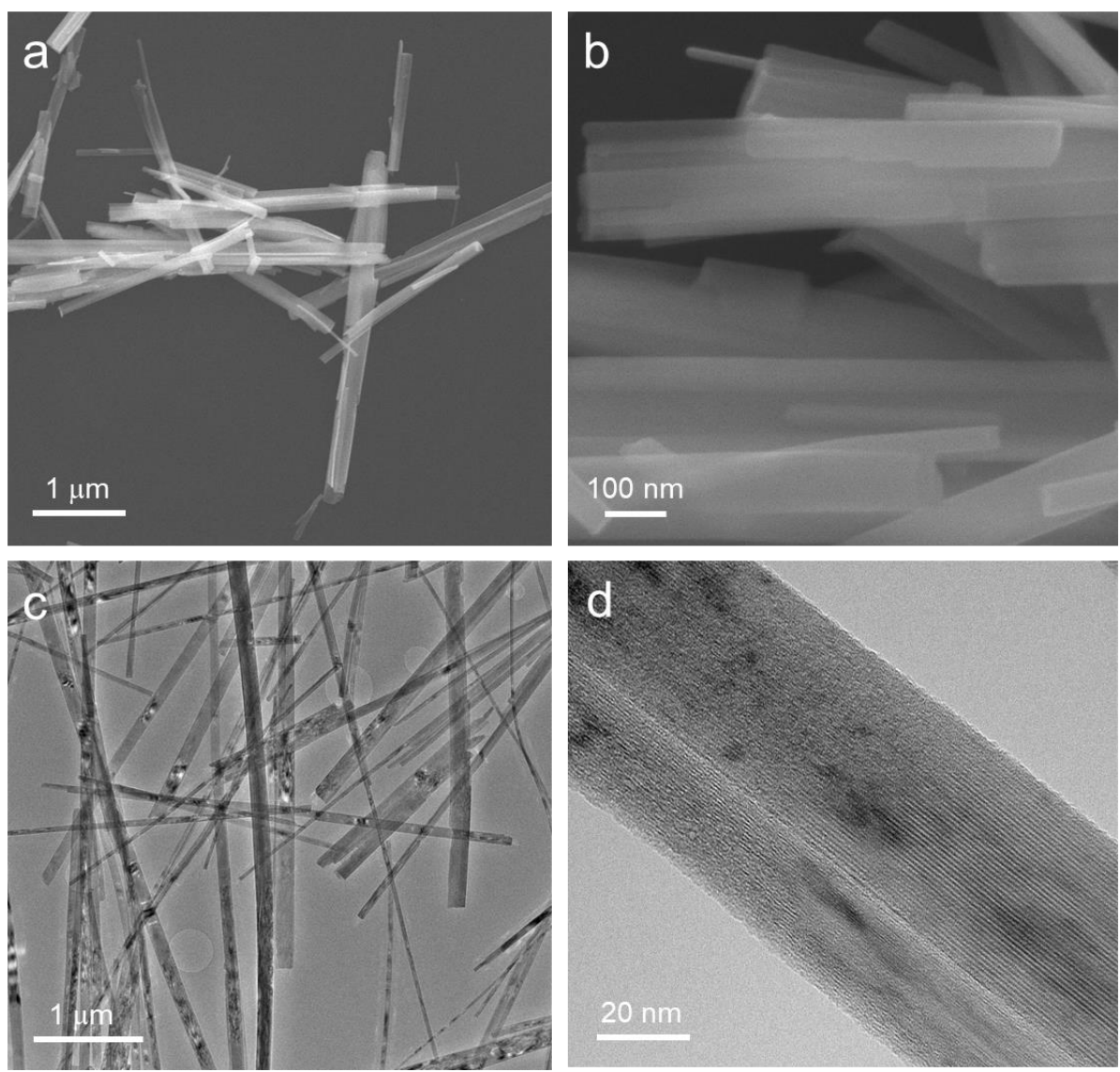

**Fig. S1 Morphology of hydrogen titanate nanowires ($H_2Ti_3O_7$).** (a, b) SEM images and (c, d) TEM images of the as synthesized hydrogen titanate nanowires ($H_2Ti_3O_7$).

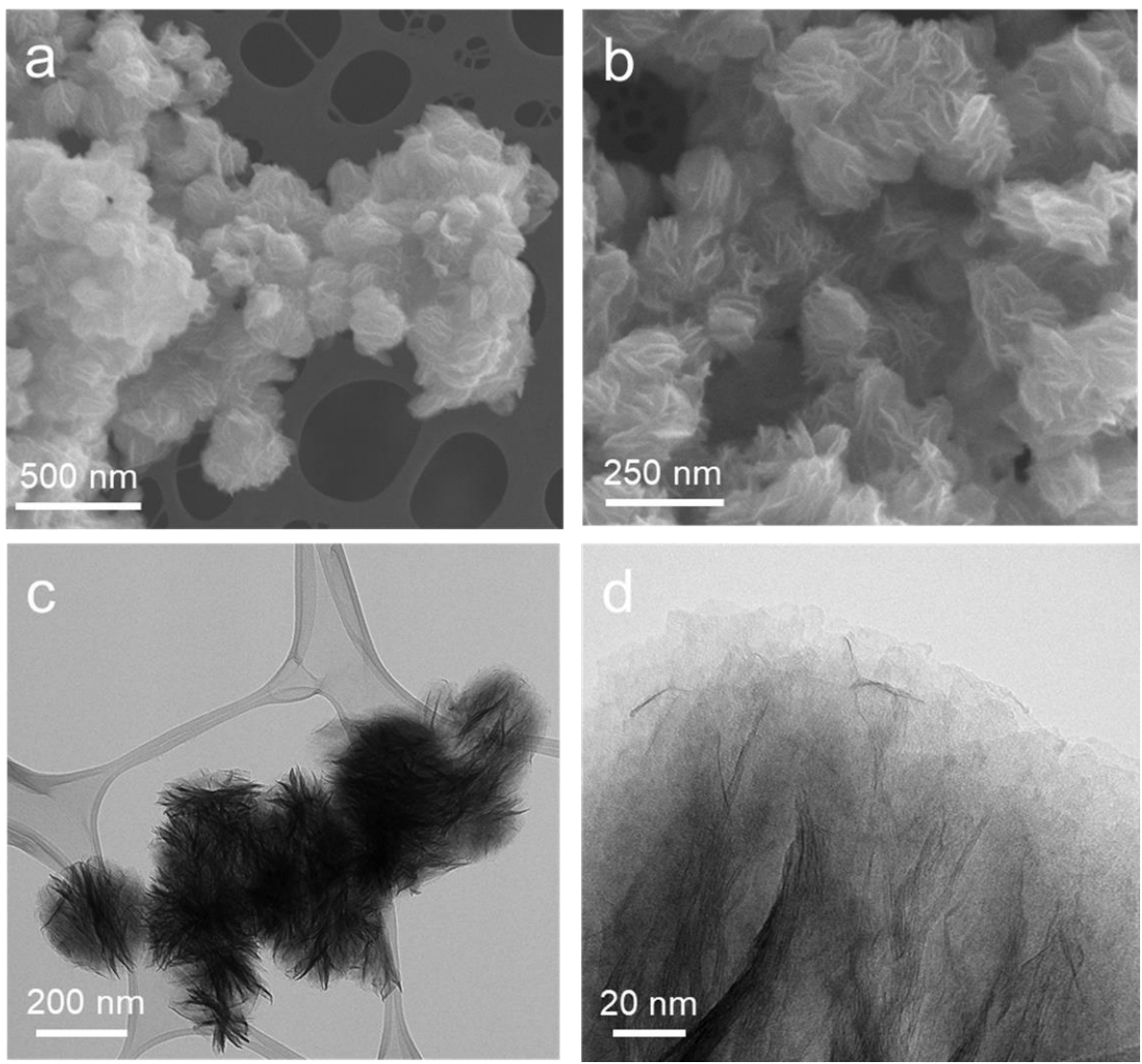

**Fig. S2 Morphology of TiO$_2$ nanospheres.** (a, b) SEM images and (c, d) TEM images of the as synthesised TiO$_2$ bronze phase nanospheres. These nanostructures are composed by spherical 3D superstructures decorated with ultrathin edges due to the assembly of ultrathin TiO$_2$ nanosheets.

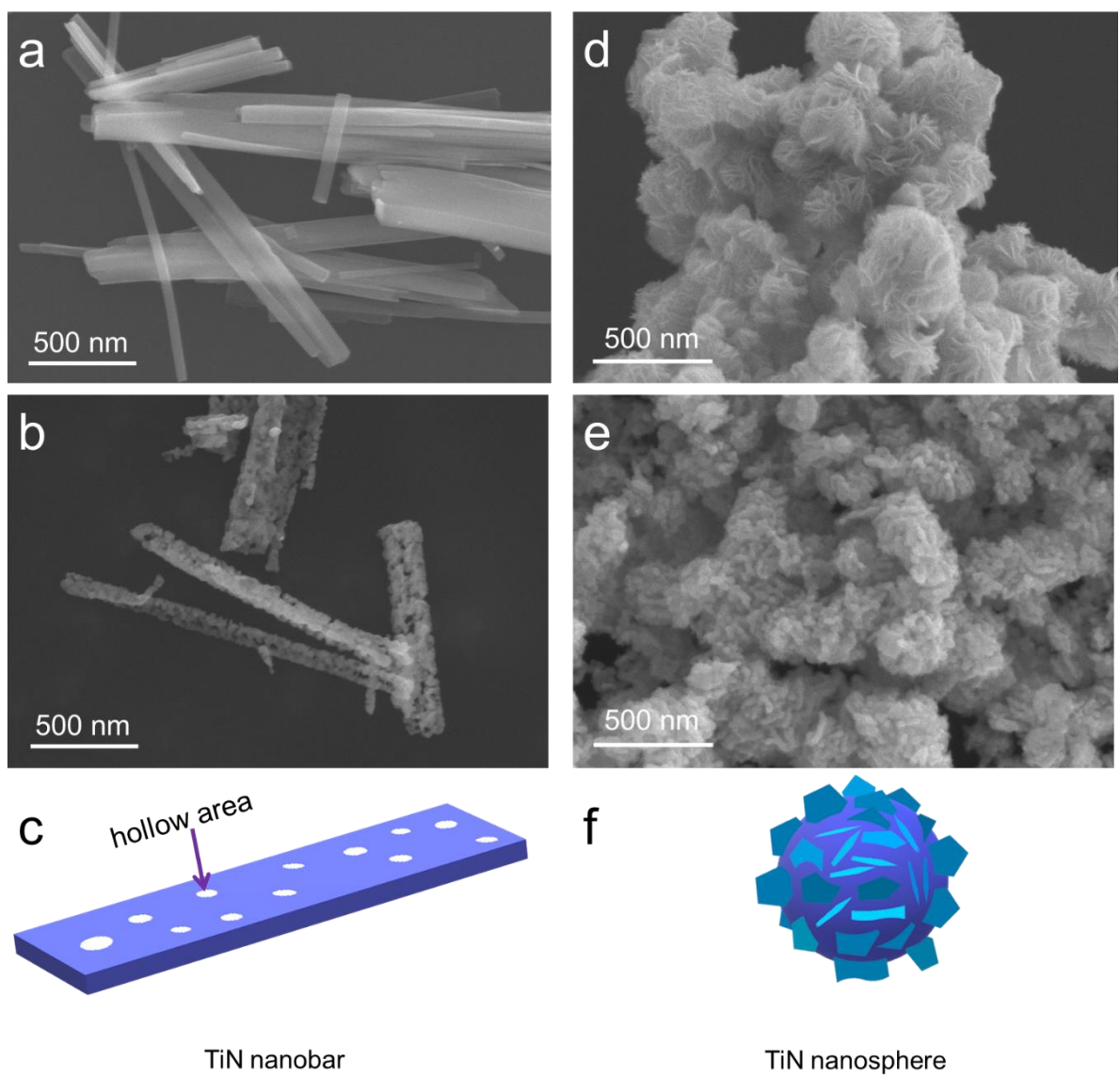

**Fig. S3** *Pseudomorphic conversion* **of TiO$_2$ nanostructures to TiN *via* high temperature nitridation.** (a) SEM image of TiO$_2$ nanowires obtained after calcination of previously synthesized hydrogen titanate nanowires (H$_2$Ti$_3$O$_7$) at 400 °C in air. (b) As obtained TiN nanobars via nitridation at 800 °C for 5 h. (c) Schematic representation of TiN nanobars. (d) SEM image of TiO$_2$ nanospheres obtained after HCl treatment of previously synthesized TiO$_2$-bronze nanospheres followed by calcination in air at 350 °C. (e) TiN nanospheres obtained via nitridation at 800 °C for 5 h. (f) Schematic representation of TiN nanospheres represented as spherical 3D superstructure decorated with ultrathin edges.

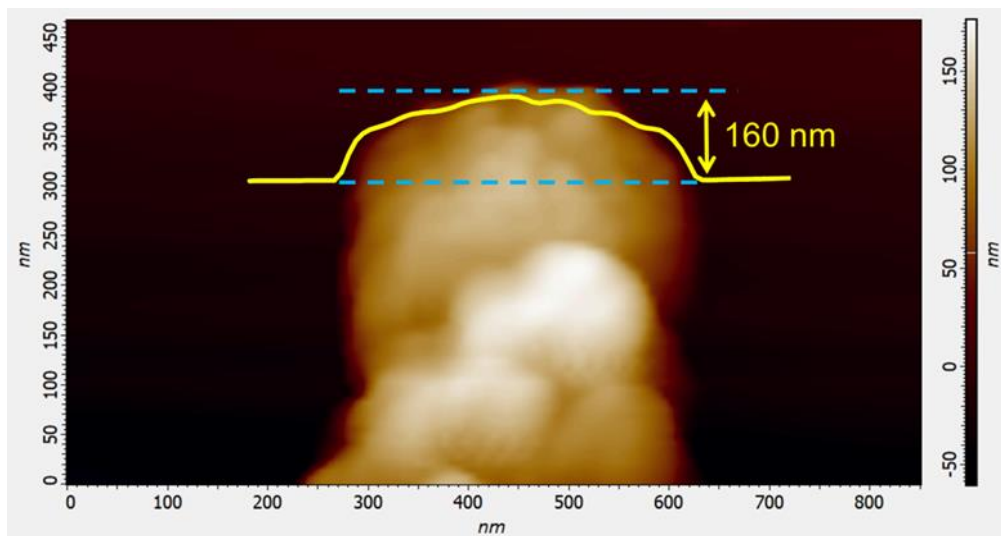

**Fig. S4 Atomic force microscopy (AFM) characterization.** AFM images and height profiles of TiN nanospheres (height ~160 nm) deposited on silicon wafer.

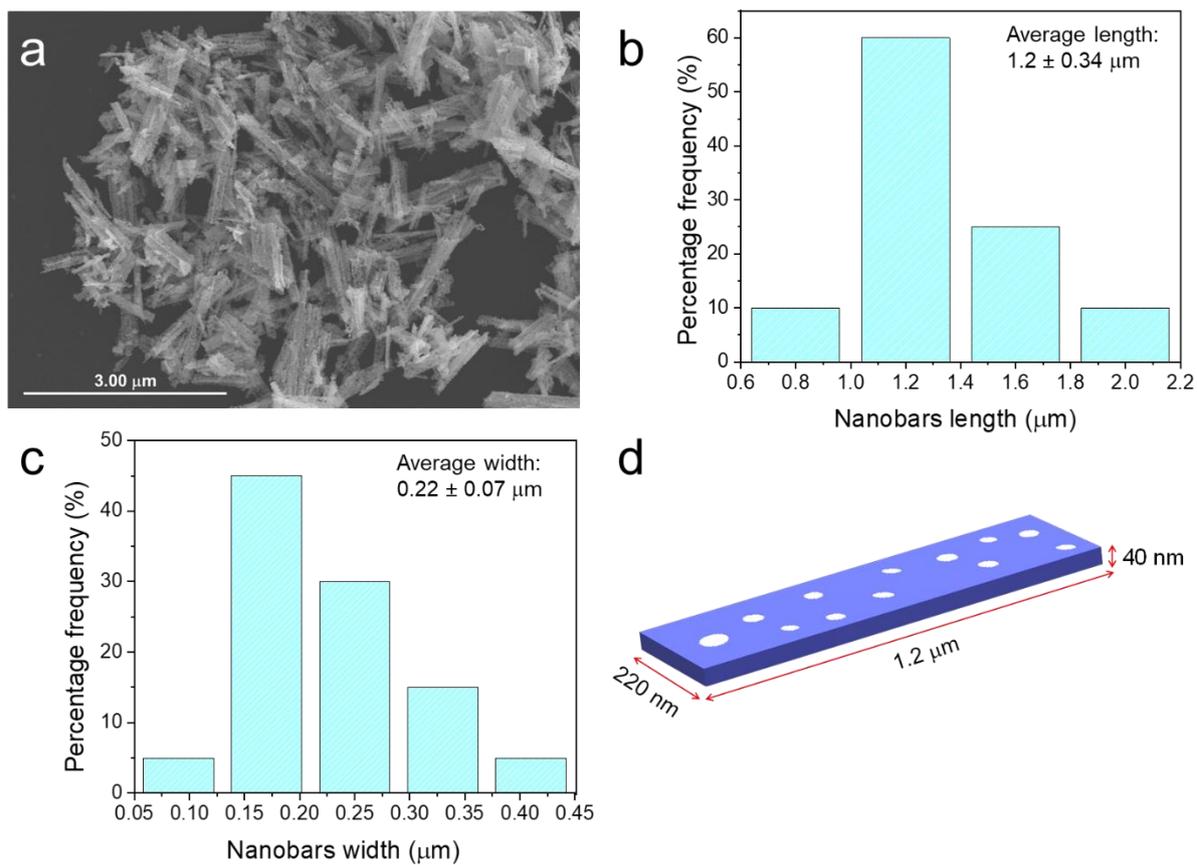

**Fig. S5 Dimensional analysis and model of TiN nanobars.** (a) Large area SEM image of TiN nanobars. Distribution of average (b) length and (c) width of TiN nanobars obtained from 100 measurements. (d) Schematics of the structure for TiN nanobars showing the pores as white circles.

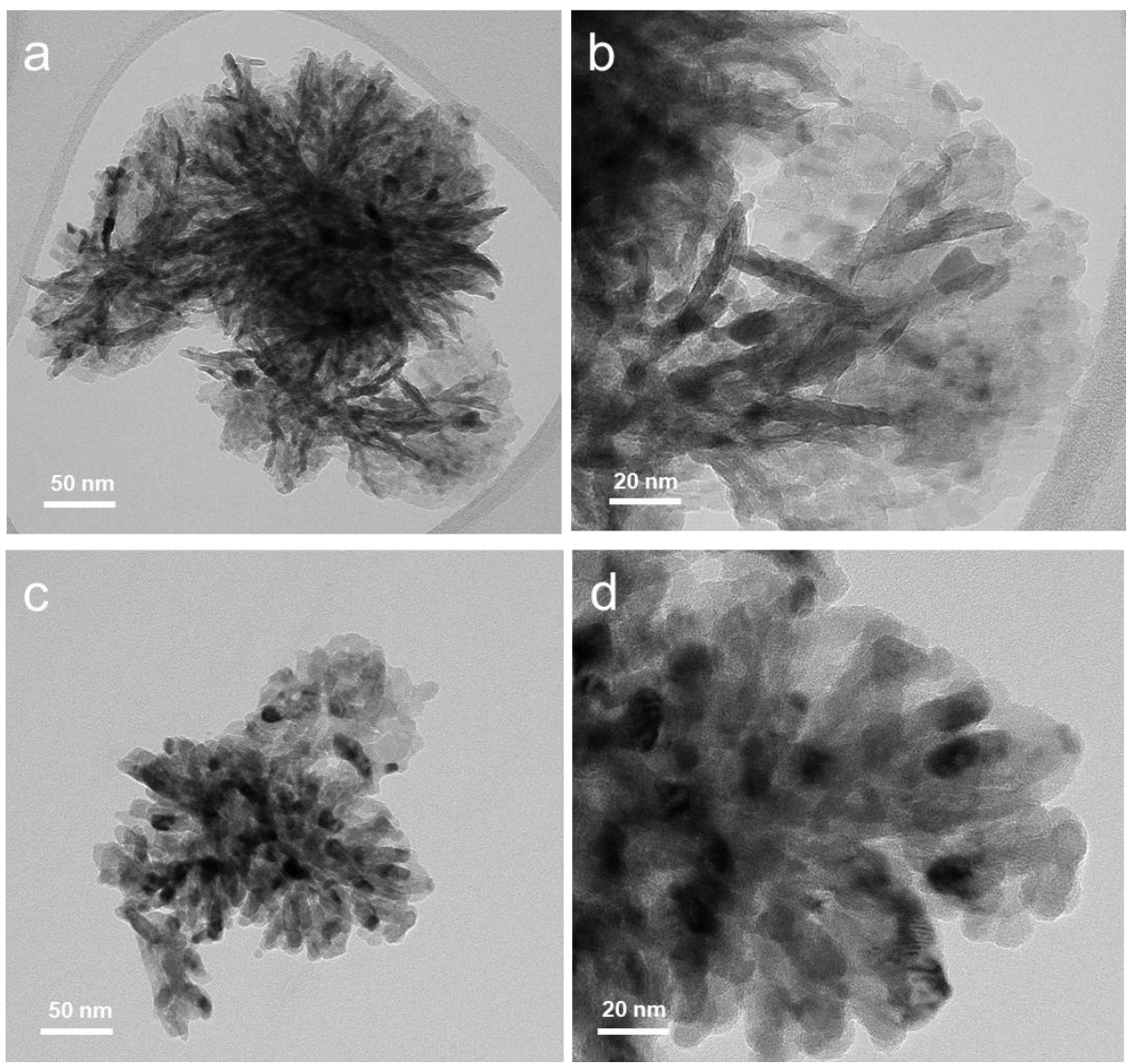

**Fig. S6 TEM characterization of TiN nanospheres.** (a) Large area and (b) magnified TEM images of TiO$_2$ nanospheres calcined at 350 °C in air. These images highlight that the 3D superstructure derives from the assembly of ultrathin nanosheets and features ultrathin edges. (c) Large area and (d) magnified TEM images of TiN nanospheres.

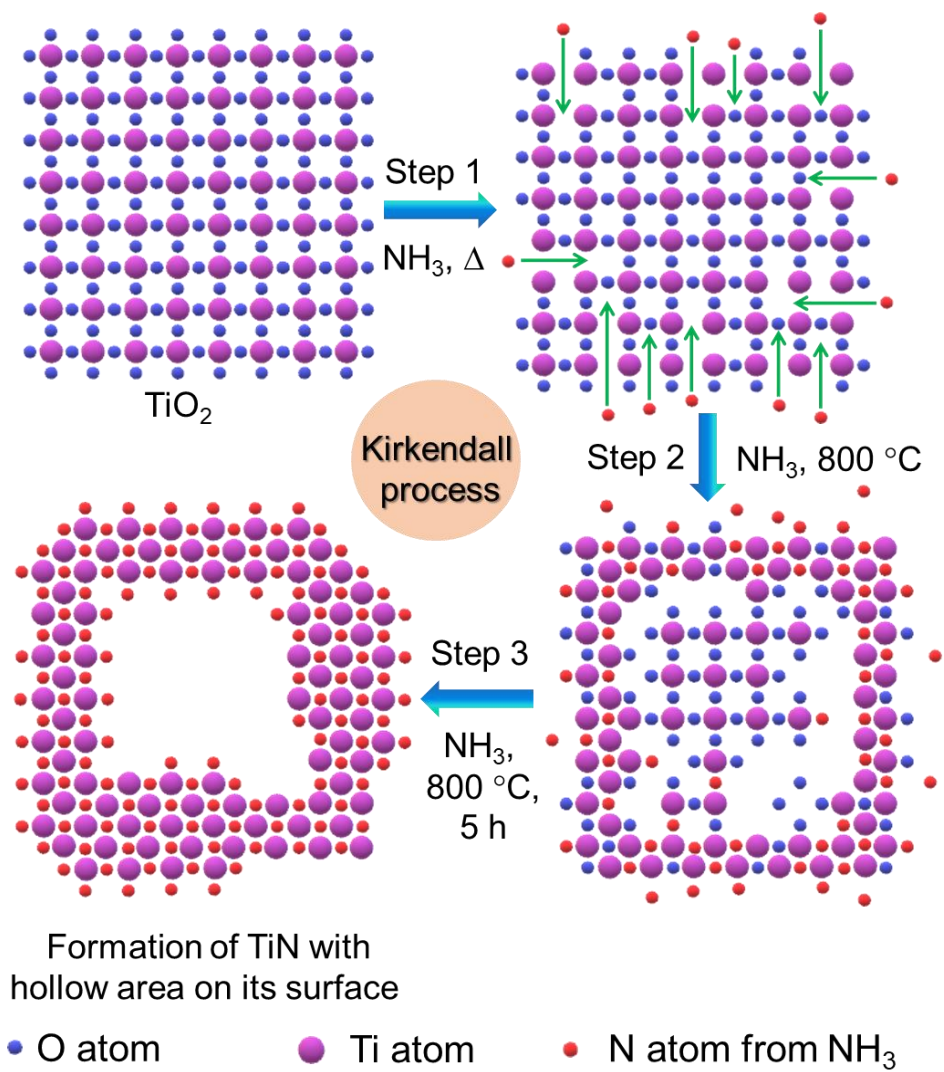

**Fig. S7 Schematic diagram for *Pseudomorphic conversion* of TiO$_2$ nanostructures to TiN via nitridation.** Proposed mechanism for the pseudomorphic conversion of TiO$_2$ nanostructures to TiN via nitridation. As shown above, a complex Kirkendall mechanism helps to understand the formation of hollow area on the surface of TiN nanobars and nanospheres structures.[S6]

The pseudomorphic conversion of TiO$_2$ to TiN via nitridation is explained and discussed in detail as follows:

1. Substitution of O atoms with N atoms into TiO$_2$ lattice requires high temperatures as it is a high energy barrier process, which often leads to inevitable aggregation, void formation via Kirkendall effect and structure modification during the reaction.

2. Stabilization of nitride anion ($N^{3-}$) in place of oxide anion ($O^{2-}$) into the lattice required one electron reduction for the conversion of $Ti^{4+}$ to $Ti^{3+}$ cation.

3. Hot $NH_3$ played a key role both as reducing agent and source of N atom.

4. Implantation of N into the $TiO_2$ lattice started from the surface (**Step 1**) with the aid of the reduction power of hot $NH_3$ and it progressively went into the interior of the lattice (**Step 2**).

5. On $TiO_2$ surface at 800 °C for 5 hrs under $NH_3$ flow, the following reactions took place leading to the formation of TiN:

$$NH_3 \leftrightarrow N + 3\,H$$

$$NH_3 \leftrightarrow N^{3-} + 3\,H^+$$

$$O^{2-} + 2\,H^+ \rightarrow H_2O_g \uparrow$$

$$Ti^{4+} + H \rightarrow Ti^{3+} + H^+$$

$$Ti^{3+} + N^{3-} \rightarrow TiN$$

6. At high temperature such drastic bond breaking and simultaneous new bond formation creates new lattice arrangements in conjugation with preserving overall starting nanostructure morphology. This is possible only because of the initial $TiO_2$ nanostructures have micron scale structural features. Any other nanoscale structures (size 10~80 nm) may undergo complete destruction of shape, size and results in fused morphology.

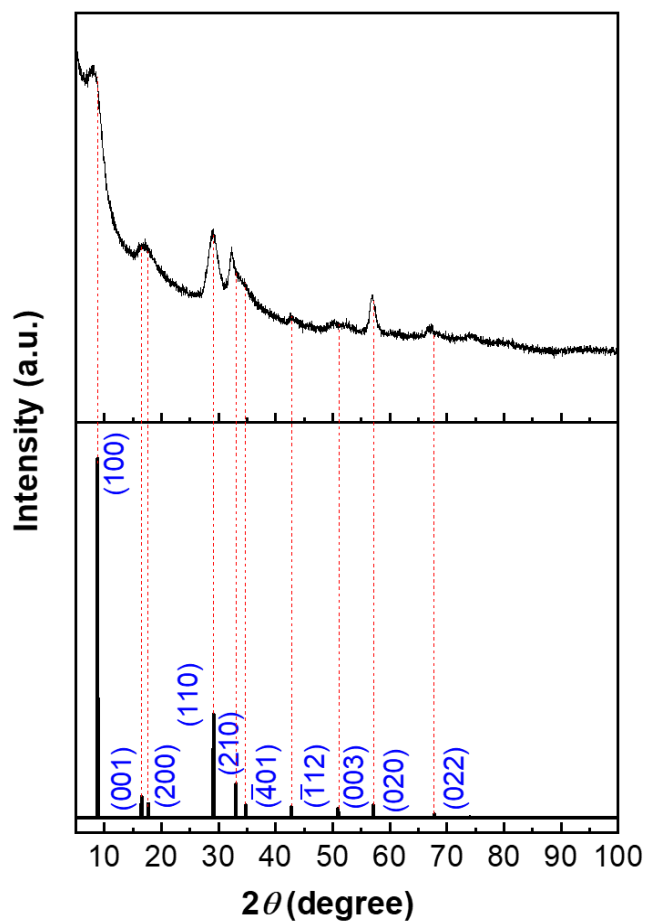

**Fig. S8 Crystal structure of TiO$_2$ nanospheres.** X-ray diffraction (XRD) patterns of TiO$_2$ nanospheres measured using Co K(α) radiation. Reference pattern of TiO$_2$ (JCPDS card no. 03-065-1156, P21/m space group, monoclinic crystal system). The diffraction patterns of TiO$_2$ nanospheres show a good agreement with the reference pattern, confirming a typical TiO$_2$−Bronze phase.

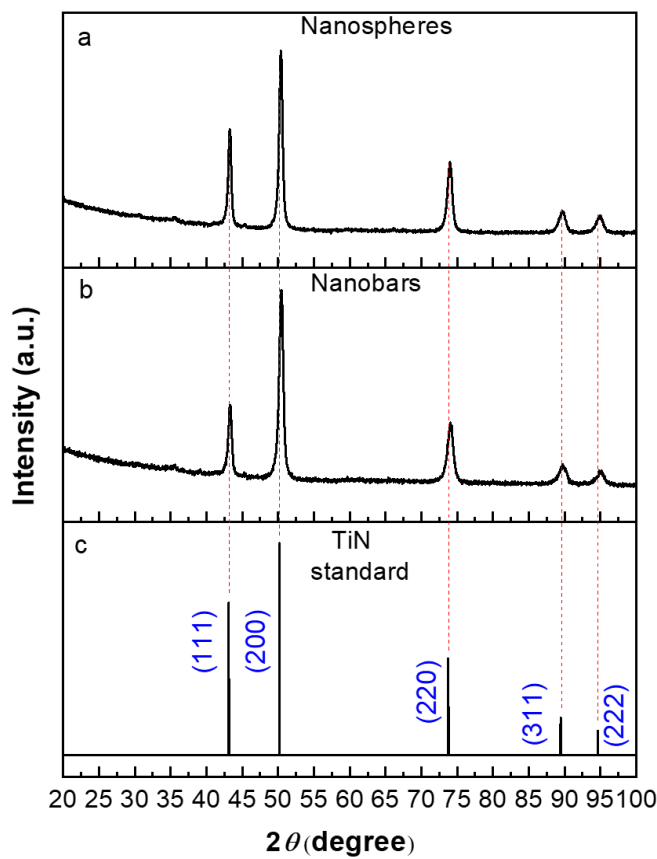

**Fig. S9 Crystal structure of TiN nanocrystals.** X-ray diffraction (XRD) patterns of (a) TiN nanospheres and (b) TiN nanobars measured using Co K(α) radiation. (c) Reference pattern of TiN (JCPDS card no. 04-015-2441, Fm-3m space group, cubic crystal system). The diffraction patterns of both TiN nanostructures show a good agreement with the reference pattern, confirming their high crystallinity. Any possible trace of $TiO_2$ or TiON phases were not detected by XRD, indicating the crystal phase purity of TiN nanostructures achieved after nitridation at 800 °C for 5 hrs.

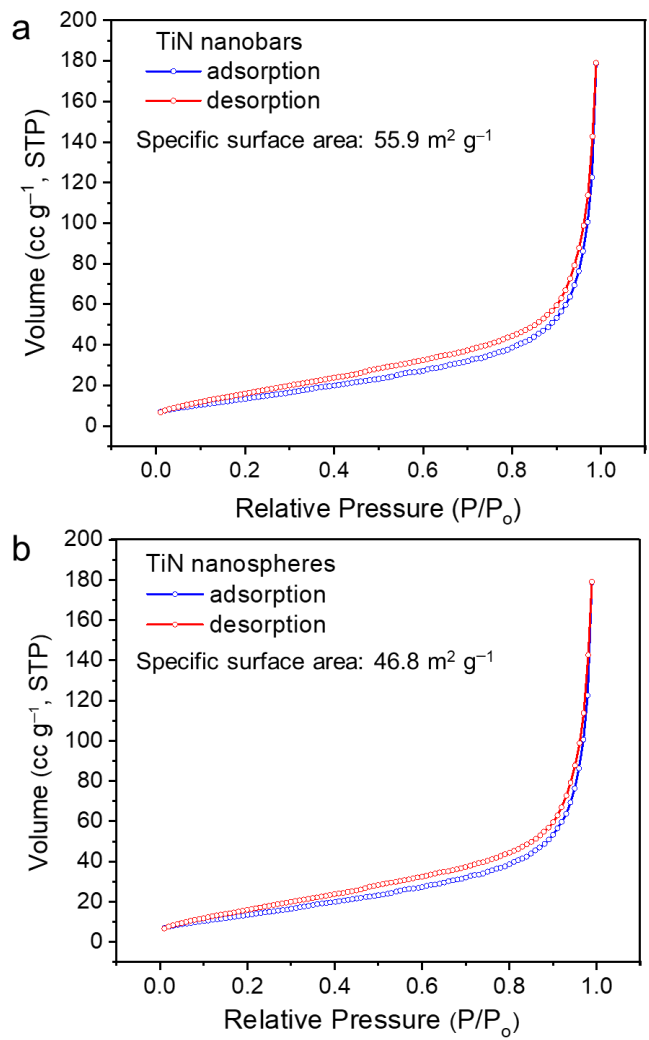

**Fig. S10 Surface area of TiN nanomaterials.** Nitrogen BET adsorption–desorption isotherms of (a) TiN nanobars and (b) TiN nanospheres. TiN nanobars have higher specific surface area (55.9 m² g⁻¹) as compared to TiN nanospheres (46.8 m² g⁻¹).

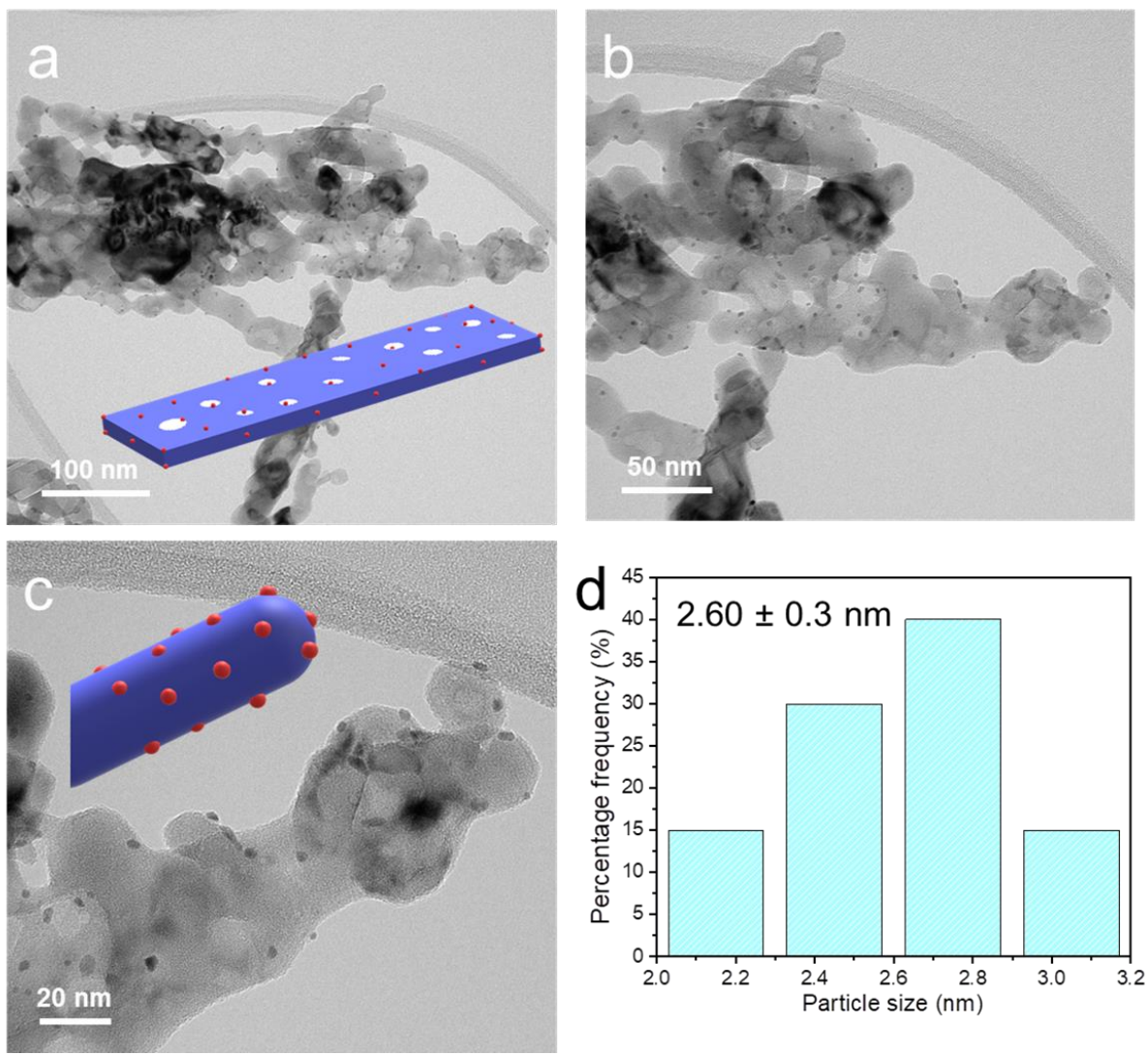

**Fig. S11 Morphological characterization of TiN/Pt nanobars.** (a–c) TEM images of TiN/Pt nanobar hybrids at three different magnifications along with schematic representations. (d) Size distribution histogram of Pt nanocrystals deposited on TiN nanobars obtained from 100 measurements.

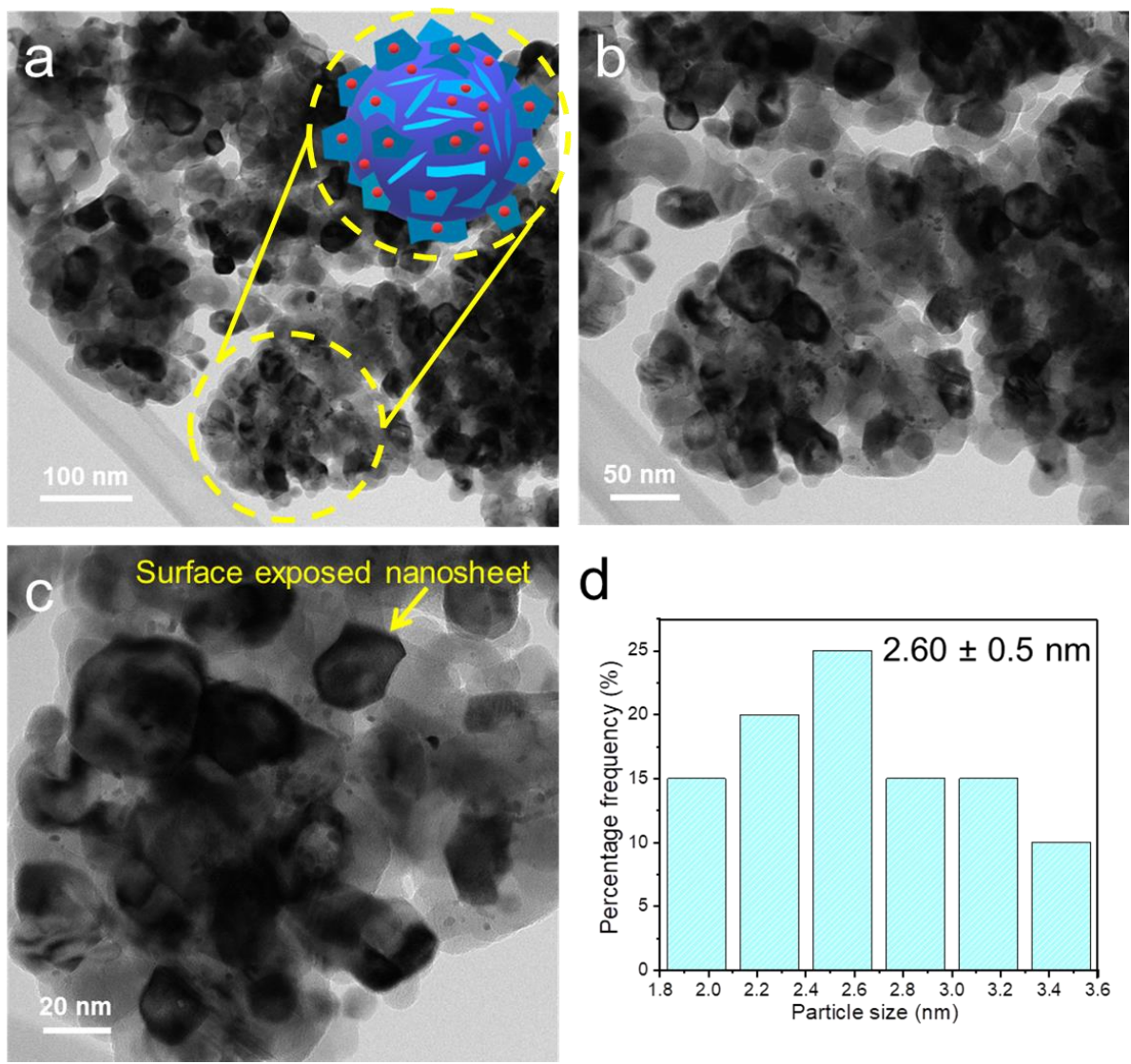

**Fig. S12 Morphological characterization of TiN/Pt nanospheres.** (a–c) TEM images of TiN/Pt nanospheres hybrid at three different magnifications with schematic representation. (d) Size distribution histogram of Pt nanocrystals deposited on TiN nanospheres obtained from 100 measurements.

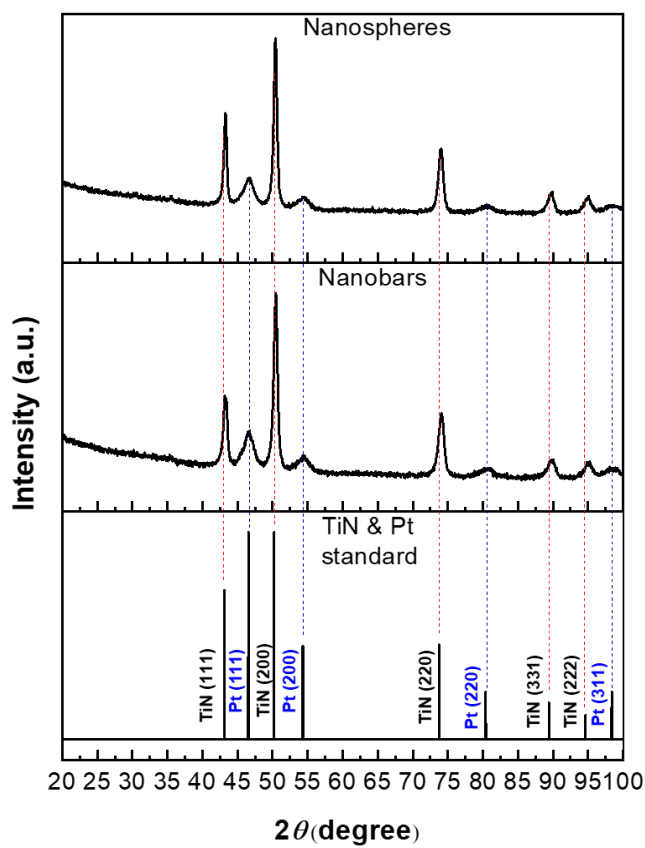

**Fig. S13 Crystal structure of hybrid TiN/Pt nanocrystals.** X-ray diffraction (XRD) patterns of (a) TiN/Pt nanospheres and (b) TiN/Pt nanobars measured using Co K(α) radiation. (c) Reference patterns for TiN (JCPDS card no. 04-015-2441, Fm-3m space group, cubic crystal system) and Pt (JCPDS card no. 01-087-0646, Fm-3m space group, cubic crystal system). The analogous TiN diffraction peaks present in both patterns confirmed the high crystallinity and the good thermal stability of TiN at 200 °C under reducing conditions employed for Pt nanoparticles preparation. The diffraction patterns for TiN/Pt nanohybrids revealed the crystalline nature of both TiN and Pt nanostructures.

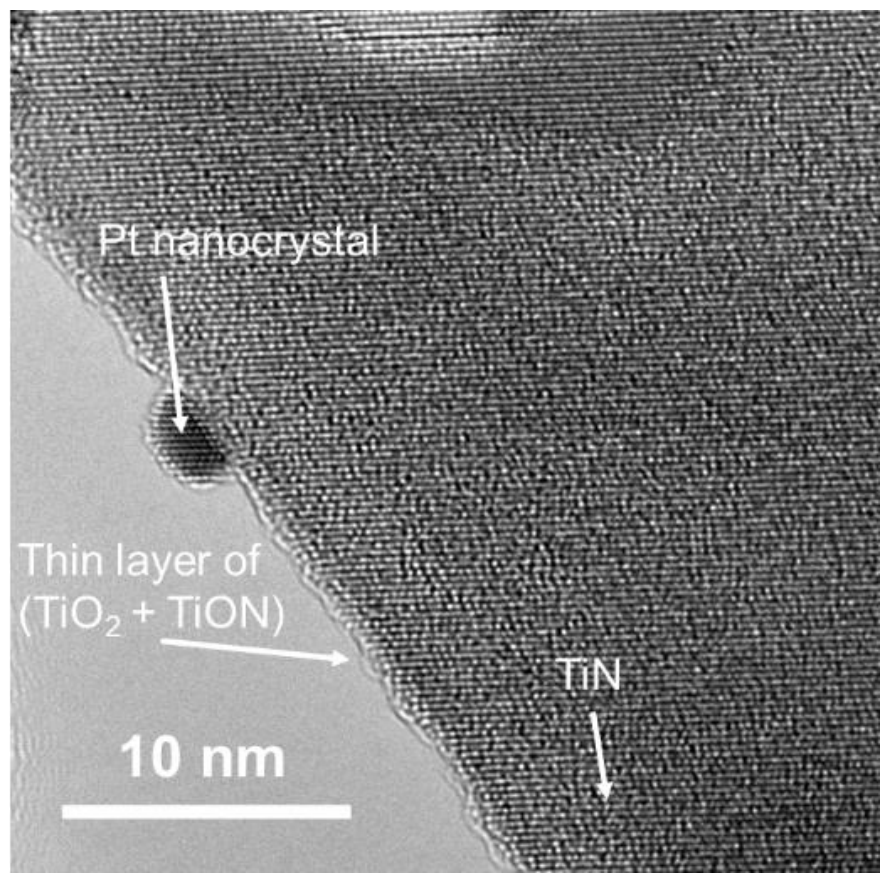

**Fig. S14 High-resolution TEM of TiN/Pt nanobars surface morphology.** Representative HRTEM image of TiN/Pt showing the thin surface passivation layer composed of $TiO_2$ and TiON. This amorphous layer has a thickness of 0.5–1 nm, in agreement with the EDS oxygen mapping reported in the main text (Fig. 1i).

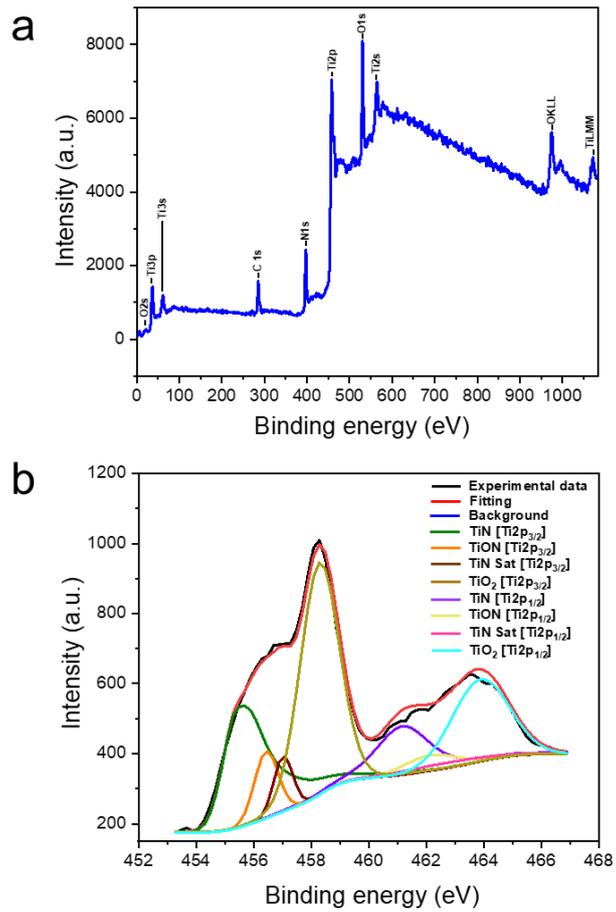

**Fig. S15 Surface characterization of TiN nanobars.** (a) Survey XPS scan of TiN nanobars. (b) High resolution XPS spectrum of the Ti 2p region with corresponding deconvolution showing that, as expected, the TiN surface is partially oxidized featuring the presence of $TiO_2$ and TiON.

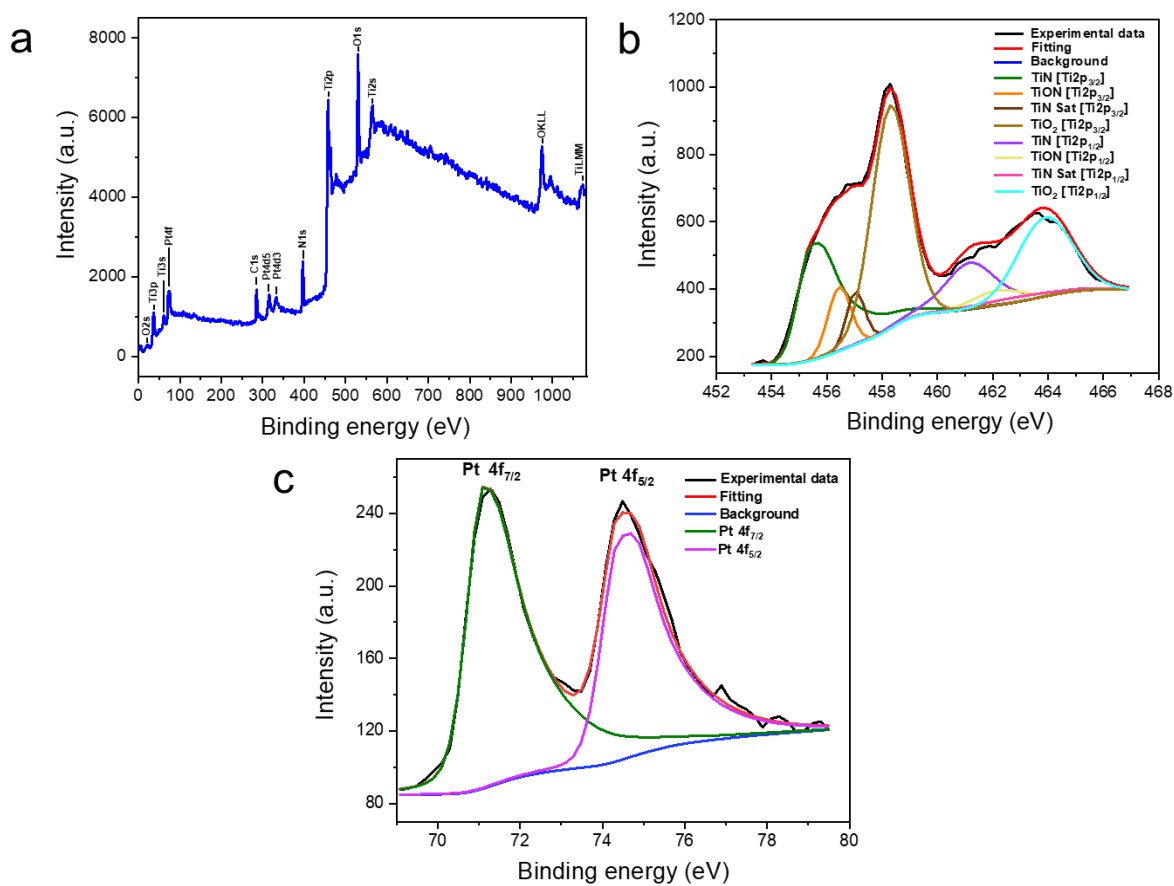

**Fig. S16 Surface characterization of TiN/Pt nanobars.** (a) Survey XPS scan of TiN/Pt nanobars. (b) High resolution XPS spectrum of the Ti 2p region with corresponding deconvolution showing that, as expected, the TiN surface is partially oxidized featuring the presence of $TiO_2$ and TiON. (c) High resolution XPS spectrum of the Pt 4f region with corresponding deconvolution confirming that Pt nanoparticles featured the $Pt^0$ oxidation state.

**Table S1. Comparison of surface layer composition in different plasmonic materials.** The relative amount (at./at. %) of each Ti species was calculated by using the area percentage retrieved from XPS spectra fitting. The amount of $TiO_2$, TiON and TiN species are compared with hybrid TiN-Pt system prepared with commercially available TiN nanocubes.[31] Notably, the surface composition of the TiN nanobars showed a lower oxide amount as compared to the commercial TiN nanocubes, denoting a higher resistance to oxidation.

| Composition | TiN nanobars | TiN/Pt nanobars | TiN/Pt nanocubes[31] |
|---|---|---|---|
| $\dfrac{TiO_2}{TiO_2 + TiON + TiN}$ | 0.4350 | 0.5157 | 0.543 |
| $\dfrac{TiON}{TiO_2 + TiON + TiN}$ | 0.0988 | 0.0879 | 0.156 |
| $\dfrac{TiN}{TiO_2 + TiON + TiN}$ | 0.4662 | 0.3964 | 0.299 |

**Optical Properties and Theoretical Simulations**

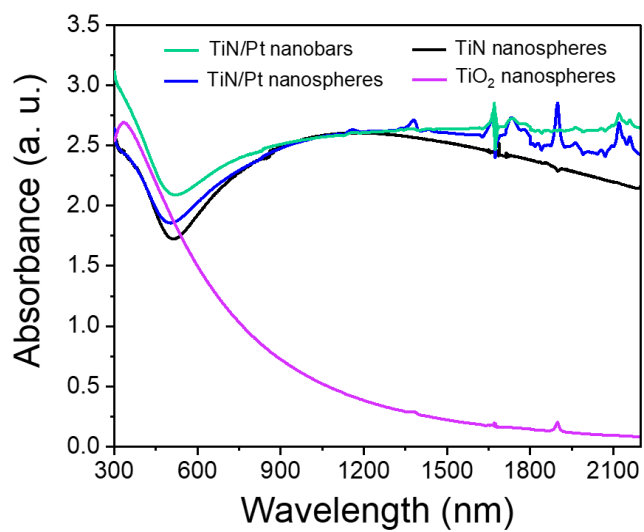

**Fig. S17 Absorption spectra.** UV-vis-NIR absorption spectra of $TiO_2$, TiN and TiN/Pt nanospheres and TiN/Pt nanobars in dichloromethane.

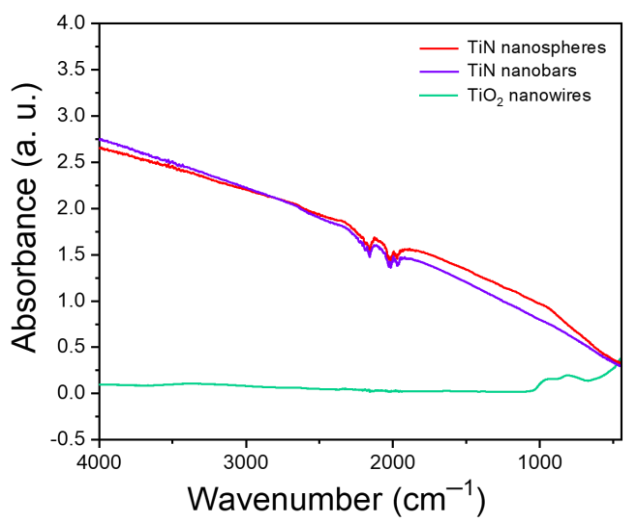

**Fig. S18 Near- (NIR), mid- (MIR), and far- infrared optical properties.** ATR-FTIR spectra of $TiO_2$ nanowires, TiN nanospheres and TiN nanobars. The spectrum of $TiO_2$ nanowires show the typical features of $TiO_2$ nanomaterials. In the case of TiN nanomaterials, the highest absorption value is observed in the NIR and MIR spectral regions (between 4000 and 2000 cm$^{-1}$) and it decreases at lower wavenumbers (higher wavelength).

**Table S2. Geometrical parameters of TiN nanostructures employed for electromagnetic simulations.**

| Geometry | Size (length/diameter) (nm) | Width (nm) | Thickness (nm) |
|---|---|---|---|
| TiN nanobars | 200, 400, 800, 1200 | 220 | 40 |
| TiN nanospheres | 200, 400, 800, 1200 | n/a | n/a |

**Table S3. Drude parameters for the Guler-Drude approximation, dichloromethane properties and parameters used in the effective medium theory employed for electromagnetic simulations.**

| Material | Dielectric Function | | |
|---|---|---|---|
| TiN | $\varepsilon_{b,Drude} = 8.76$ | $\omega_p = 6.94$ eV | $\Gamma_D = 0.297$ eV |
| Dichloromethane | $\varepsilon = 2.0289$ | | |

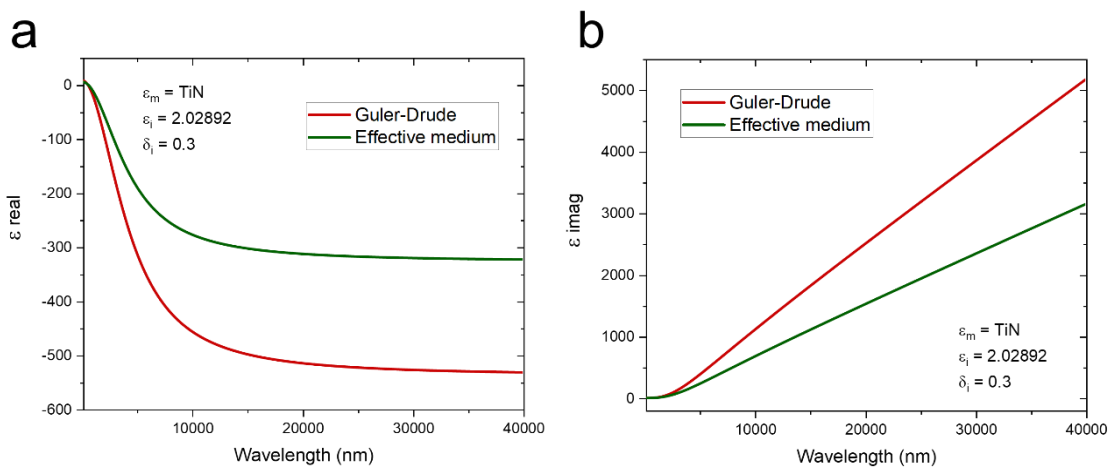

**Fig. S19 Dielectric function for TiN.** The (a) real and (b) imaginary parts of the TiN dielectric function calculated from the Drude approximation and the effective medium theory.

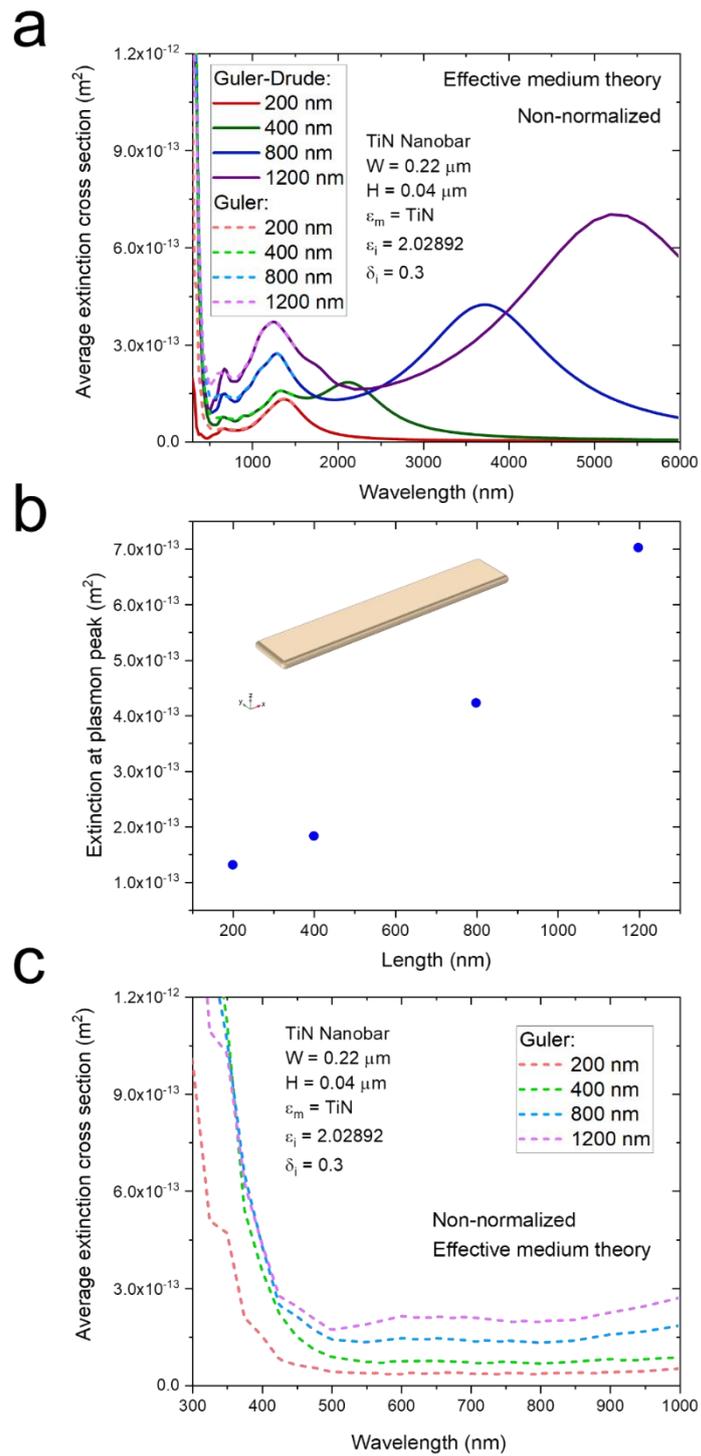

**Fig. S20 Extinction of light for the TiN nanobar computed within the effective medium model (COMSOL).** Average extinction cross section of TiN nanobars for: (a) different nanobars lengths in the UV – mid-IR range, (b) at the main plasmon resonance, and (c) in the [UV-near-IR] 300 - 1000 nm interval, calculated using the effective medium theory.

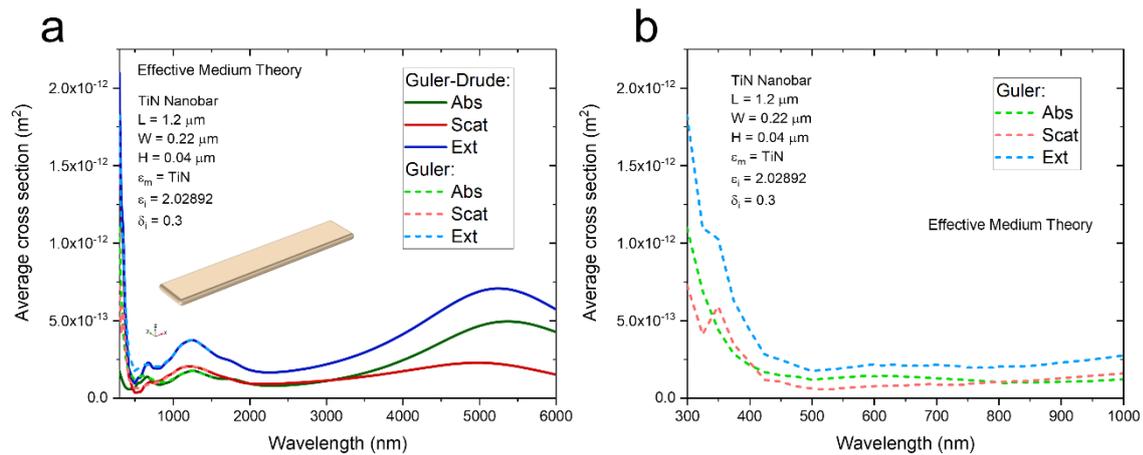

**Fig. S21 Detailed optical cross sections within the effective medium model (COMSOL).** Absorption, scattering and extinction cross sections averaged over six incident light directions in the intervals (a) (from UV to mid-IR) 300 – 6000 nm and (b) (from UV to near-IR wavelengths) 300 -1000 nm.

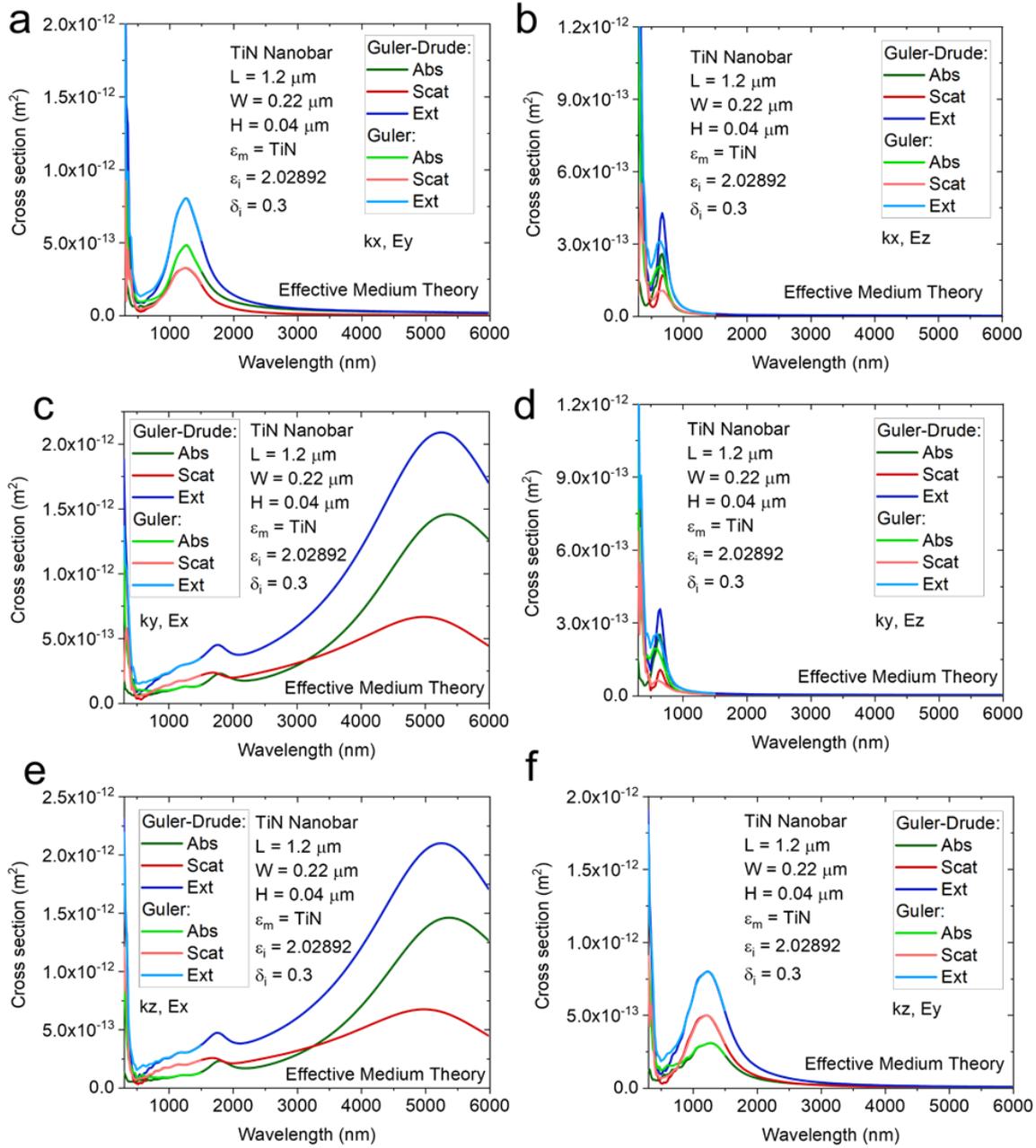

**Fig. S22 Polarized optical cross sections for the six incident light directions, within the effective medium model (COMSOL).** Optical cross sections for TiN nanobars where each subplot corresponds to one incident light direction (three wavevector directions and two polarizations). Every wavevector direction excites a different mode (one longitudinal and two transversal).

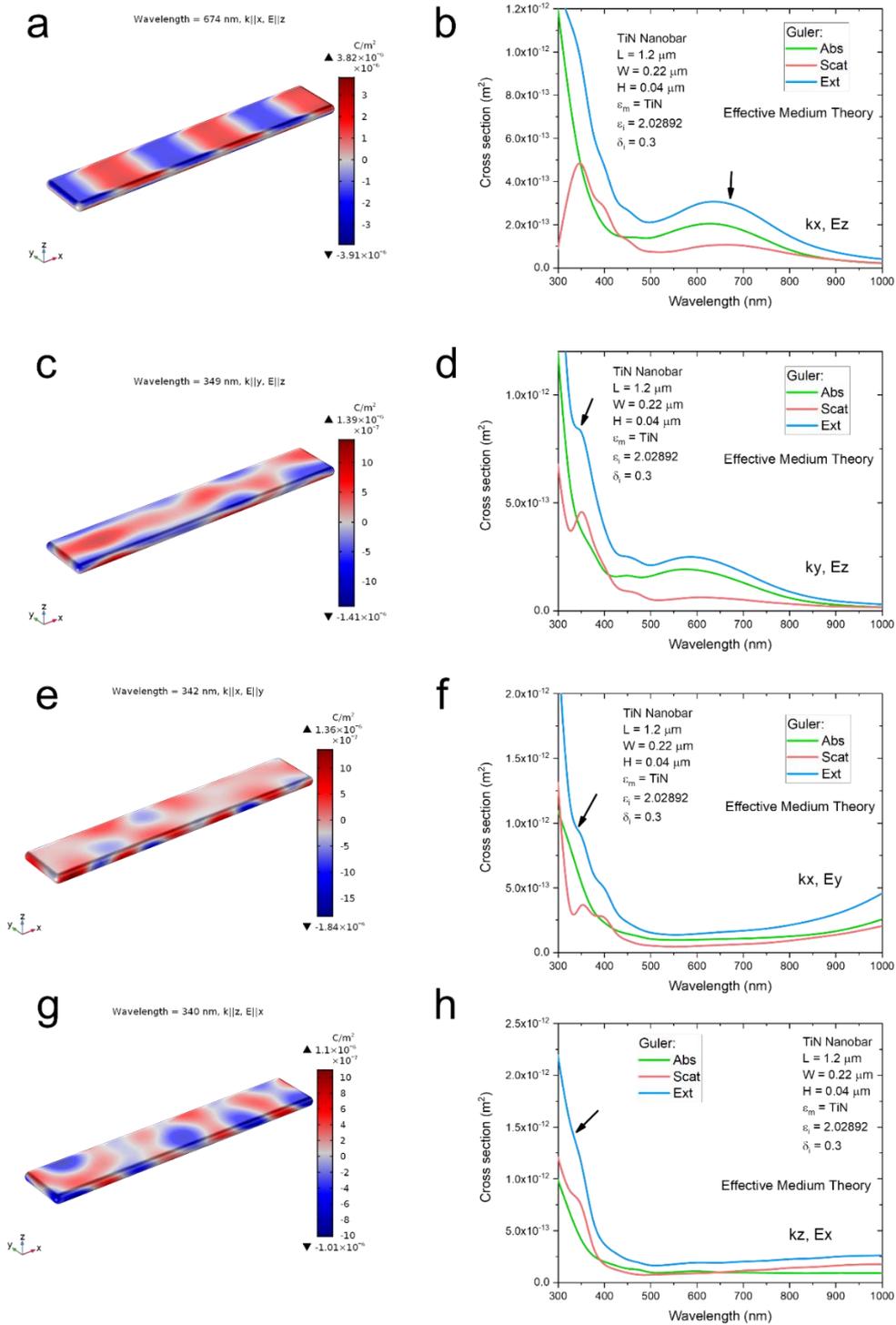

**Fig. S23 Ultraviolet and visible modes within the effective medium model (COMSOL).** Surface charge maps (a, c, e, g) and optical cross sections (b, d, f, h) corresponding to a specific direction and polarization respectively. The surface charge maps correspond to a specific wavelength, signaled by an arrow on the cross-section plots.

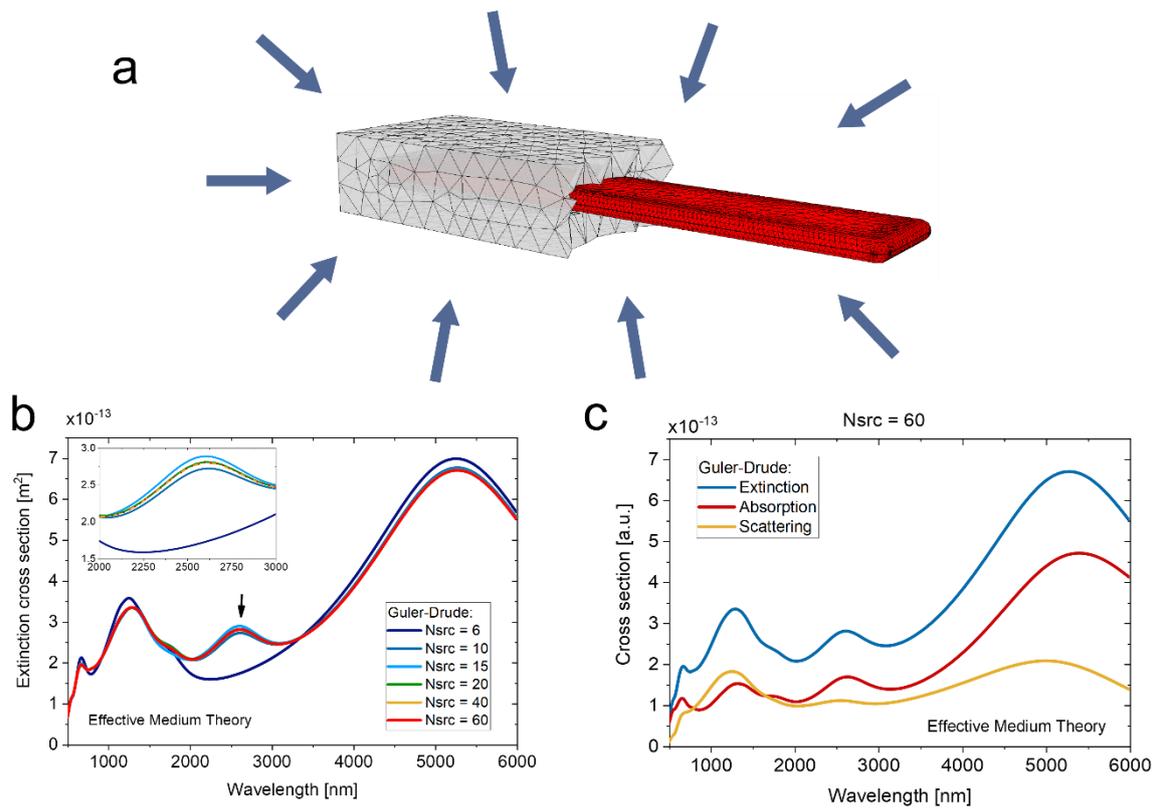

**Fig. S24 Cross section for TiN nanobar simulation (JCMsuite).** (a) TiN nanobar simulation model. (b) Extinction cross sections calculated with different numbers of incident light directions, Nsrc. (c) Scattering, absorption, and extinction cross sections calculated with 60 incident light directions. We utilized the effective medium model.

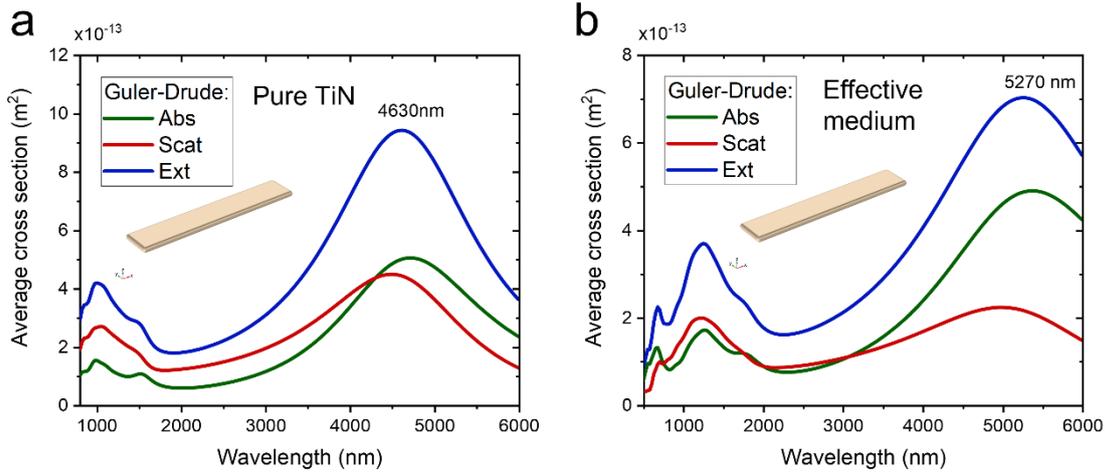

**Fig. S25 TiN nanobar vs hybrid medium nanobar (COMSOL).** Average optical cross sections for two 1200 nm-long nanobars which corresponds to (a) TiN nanobar with a Guler-Drude dielectric function, while (b) corresponds to an effective medium (dielectric function) model based on the Guler dielectric function. The main differences between the nanobars response to light is the extinction intensity and the plasmon resonance wavelength. Additionally, the absorption cross section is more prominent for an effective medium dielectric function.

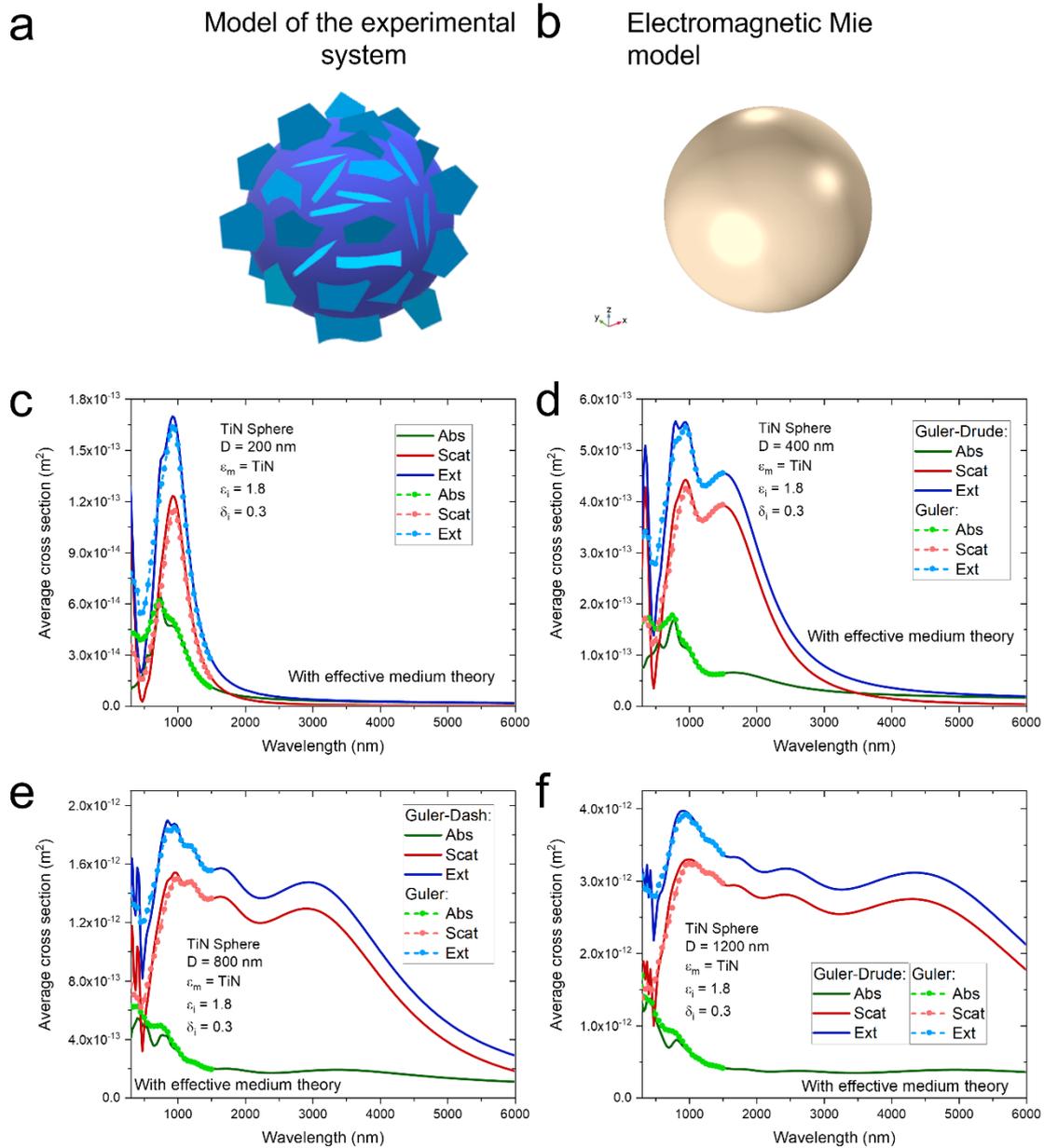

**Fig. S26 Model and average cross sections for the TiN nanospheres (COMSOL).** (a) model of nanospheres structures obtained from material characterization and (b) optimized electromagnetic Mie model used for simulation. Average cross sections for TiN nanospheres with different diameters: (c) 200 nm, (d) 400 nm, (e) 800 nm, and (f) 1200 nm. A Guler-Drude dielectric function and the original Guler empirical dielectric function were used. In both cases, the Maxwell-Garnett effective medium theory was implemented. As the nanosphere grows, the cross sections broaden.

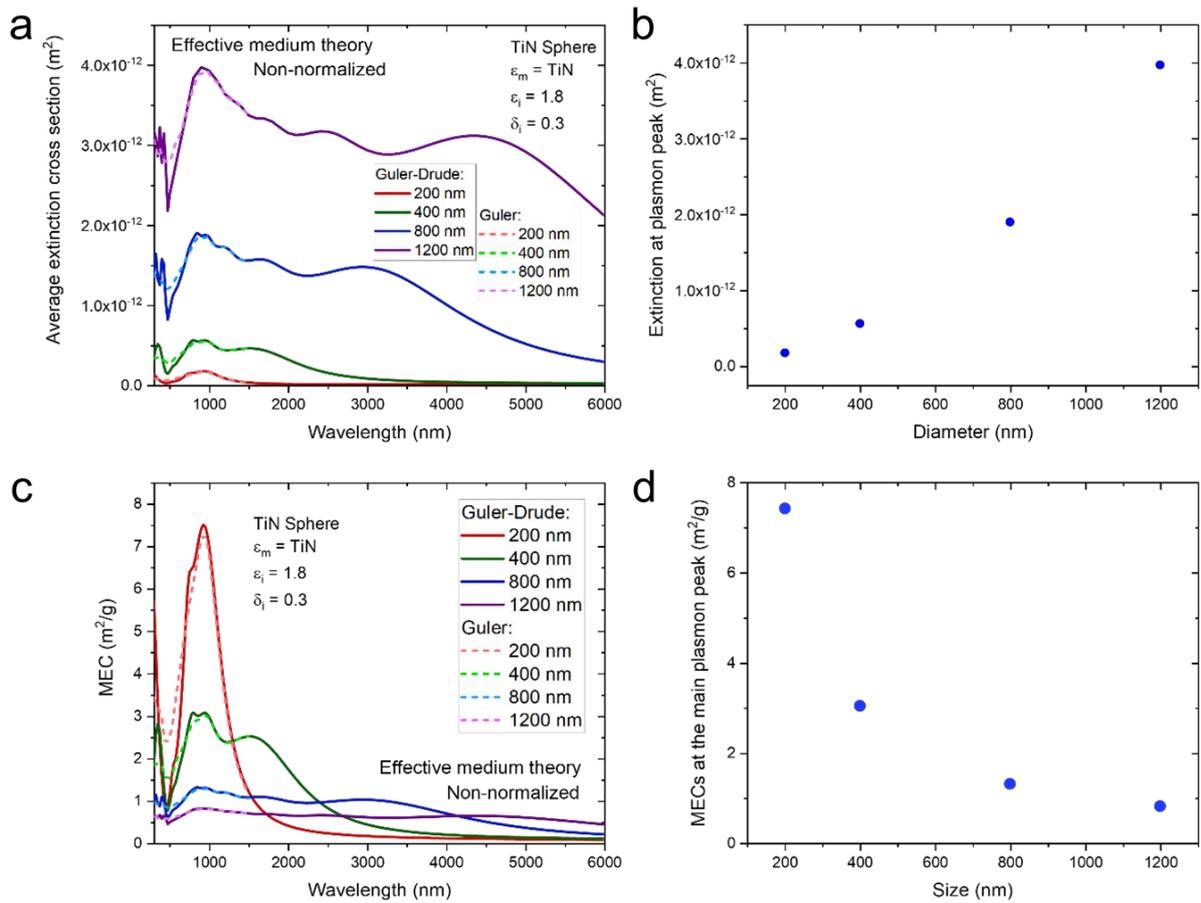

**Fig. S27 Cross sections and MEC for the TiN nanosphere (COMSOL).** (a, c) Average extinction cross section and (b, d) mass extinction coefficient (MEC) for nanospheres of different diameters such as 200, 400, 800, and 1200 nm. Both the Guler empirical dielectric function and the Guler-Drude approximation were used, as well as Maxwell-Garnett effective medium theory. The scatter plots correspond to the intensity at the main plasmon resonance.

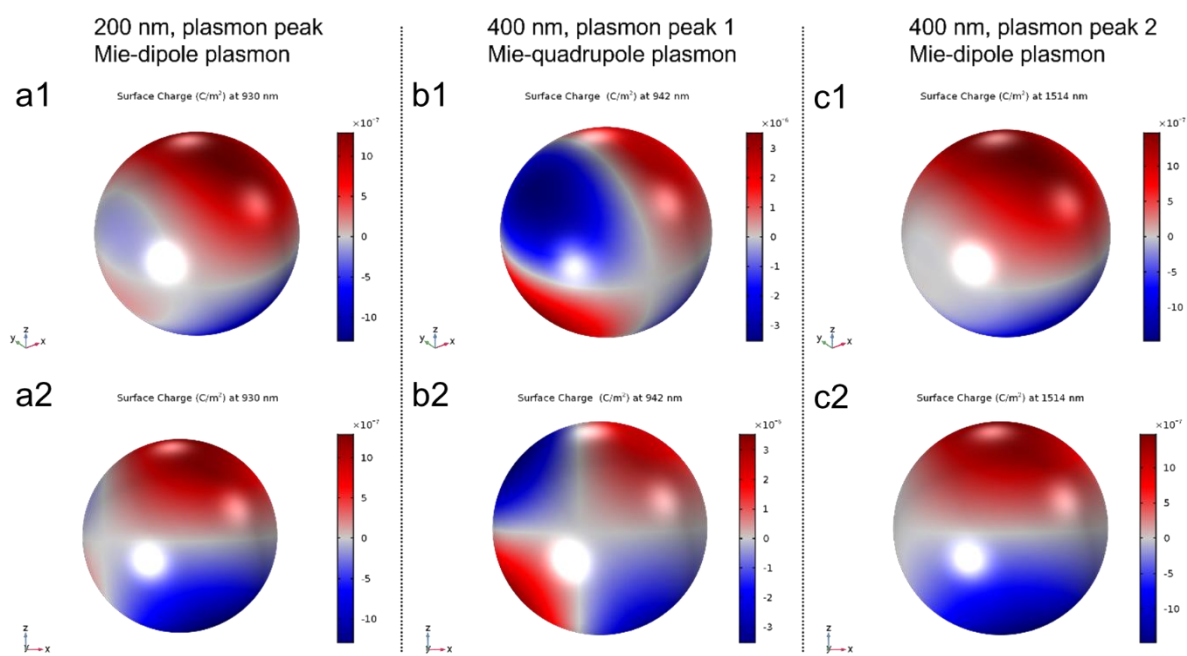

**Fig. S28 Surface charge on the TiN nanospheres (COMSOL).** Maps of the surface charge for TiN nanospheres with (a1−a2) 200 nm diameter at 930 nm, (b1−b2) 400 nm diameter at 942 nm and (c1−c2) 400 nm diameters at 1514 nm. The model: the effective medium and the Guler-Drude function. Each map is for a specific incident light direction and plasmon resonance. The 200 nm nanospheres show a dipole plasmon, while the 400 nm has two modes (dipole and quadrupole).

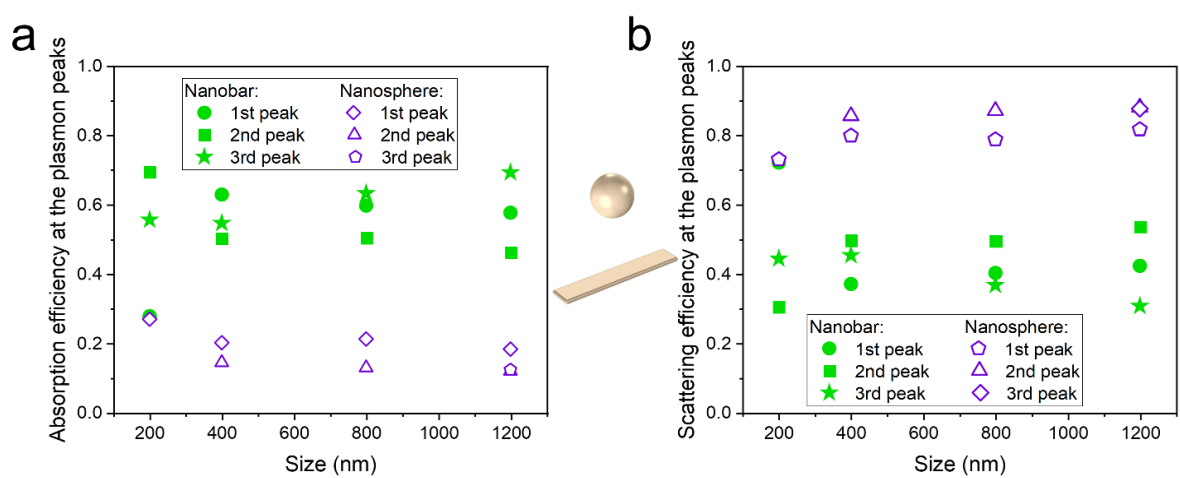

**Fig. S29** Theory for the optical efficiencies - the TiN nanorods vs. the TiN nanospheres.

**Plasmon-Enhanced Hydrogen Evolution from $NH_3BH_3$**

Turnover Frequency (TOF) Calculation

The volume of evolved $H_2$ gas with respect to time followed a linear trend and, subsequently, reached a plateau region, corresponding to the end of the reaction. The turnover frequency (TOF) on Pt catalytic sites (total amount of Pt determined from ICP measurements) was calculated considering the time needed to evolve the stoichiometric volume of $H_2$ by the decomposition of all ammonia borane molecules added to the reaction solution:

$$TOF\ (mol_{H_2}\ mol_{Pt}^{-1}\ min^{-1}) = \frac{mol_{H_2}\ generated}{mol_{Pt} \times time\ of\ reaction}$$

The general definition of TOF was then applied to find the photothermal turnover frequency, $TOF_{photo}$, and the thermal turnover frequency, $TOF_{therm}$. The former is defined as the TOF obtained under a given light intensity (H), which takes into account the contributions from both hot electrons and thermal effects upon light irradiation. The latter is defined as the TOF obtained by performing the reaction in dark at a final reaction temperature (FT, the temperature generated when the reaction was completed under light irradiation). As a result, the photocatalytic turnover frequency due to hot electrons at a specific light intensity was calculated as:

$$TOF_{hot-e}\ (H) = TOF_{photo}\ (H) - TOF_{therm}\ (H)$$

Apparent Quantum Yield (AQY) Calculation

The TOF can be re-written to express the number of $H_2$ molecules evolved per second as

$$k\left(\frac{molecules}{s}\right) = \frac{TOF \times N_A \times n_{Pt}}{60}$$

where $k$ is reaction rate constant, $N_A = 6.022 \cdot 10^{23}$ mol$^{-1}$ is the Avogadro constant, and $n_{Pt}$ is the total number of mole of Pt from ICP measurements. On the other hand, the incident photon flux ($\Phi$) was calculated as

$$\Phi = \frac{\lambda}{hc} H \left[ s^{-1} m^{-2} \right]$$

where $\lambda$ is the wavelength of incident light, $h = 6.626 \cdot 10^{-34}$ J s is the Planck constant, $c = 3 \cdot 10^{8}$ m s$^{-1}$ is the speed of light, and $H$ is the light intensity of photons (equal to 24 mW cm$^{-2}$ for all the experiments with monochromatic irradiation). By combining these two equations, the AQY was calculated as

$$AQY\% \left( \frac{molecules}{photons} \right) = \frac{k}{A \times \Phi} \times 100$$

where A is the geometric cross section of the photoreactor.

Activation Energy ($E_a$) Calculation

The activation energy ($E_a$) for the $H_2$ evolution reaction was evaluated by using the Arrhenius equation, i.e., by plotting the natural logarithm of the reaction rate (TOF in s$^{-1}$) vs. the reciprocal absolute temperature (in K). $E_a$ (dark) was calculated considering the photocatalytic rate for the reaction carried out in the dark at final temperature (FT). In contrast, $E_a$ (light) was calculated considering the photocatalytic rate at a particular light intensity (H) as follows:

$$ln\, TOF_{hot-e}\,(H) = ln\, TOF_{photo}\,(H) - ln\, TOF_{therm}\,(H)$$

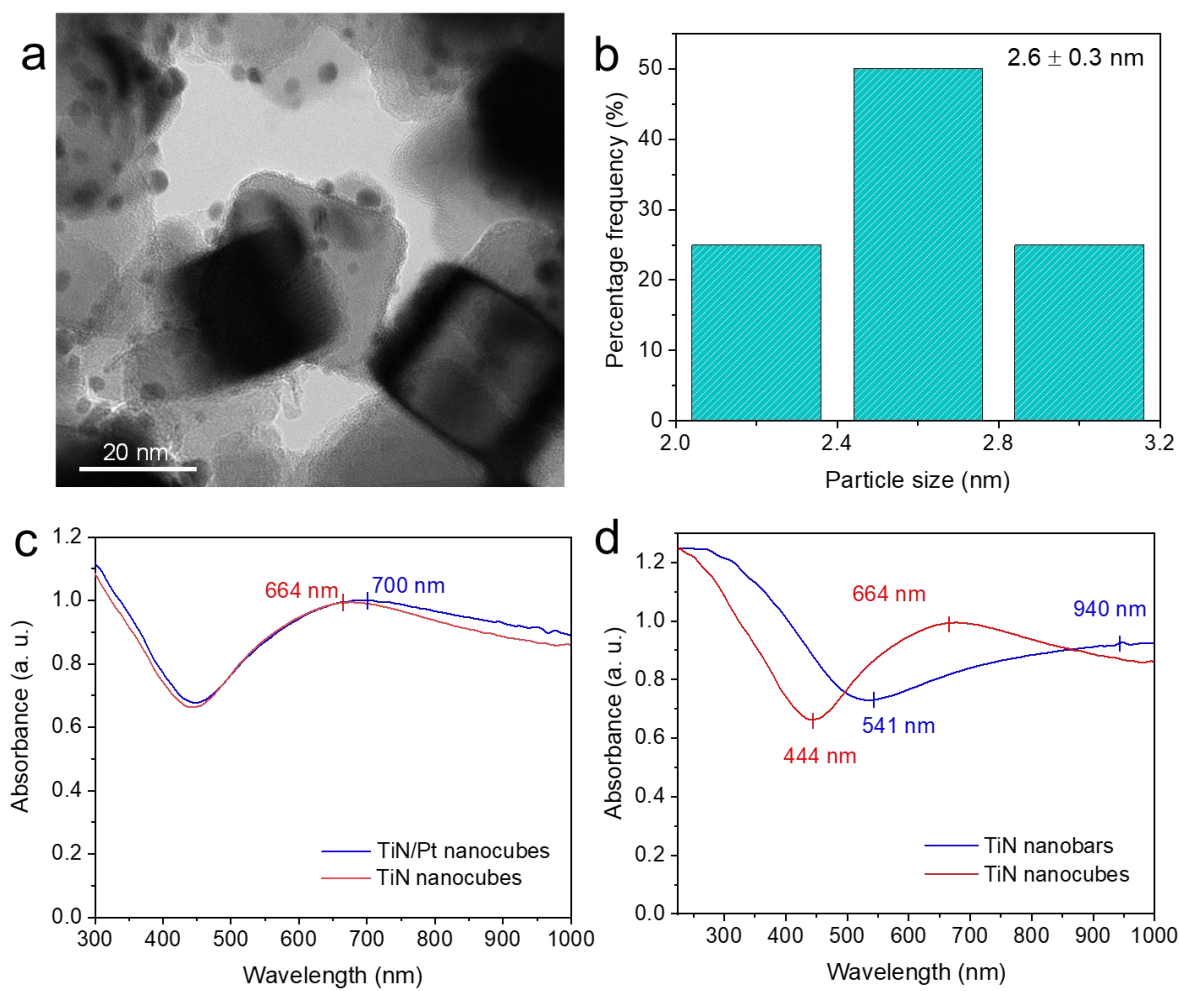

**Fig. S30 Characterization of TiN/Pt nanocubes.** (a) TEM image of TiN/Pt nanocubes. Highly disperse Pt nanocrystals can be seen on the TiN nanocubes. (b) Size distribution histogram of Pt nanocrystals deposited on TiN nanocubes obtained from 100 measurements (size: 2.6 ± 0.3 nm). (c) UV-Visible absorbance spectra of commercially available TiN nanocubes (50 nm) and TiN/Pt hybrid nanocubes in DI-$H_2O$. It shows that the decoration of TiN nanocubes with Pt nanocrystals induced a significant shift of the plasmonic peak from 664 to 700 nm as the result of the dielectric environment change at the surface of TiN because of the Pt deposition. (d) Comparison of absorbance spectra for pure commercial TiN nanocubes and as-synthesized TiN nanobars in DI-$H_2O$. Significant increase in the absorbance spectra in the NIR region can be seen for TiN nanobars as compared to nanocubes in DI-$H_2O$.

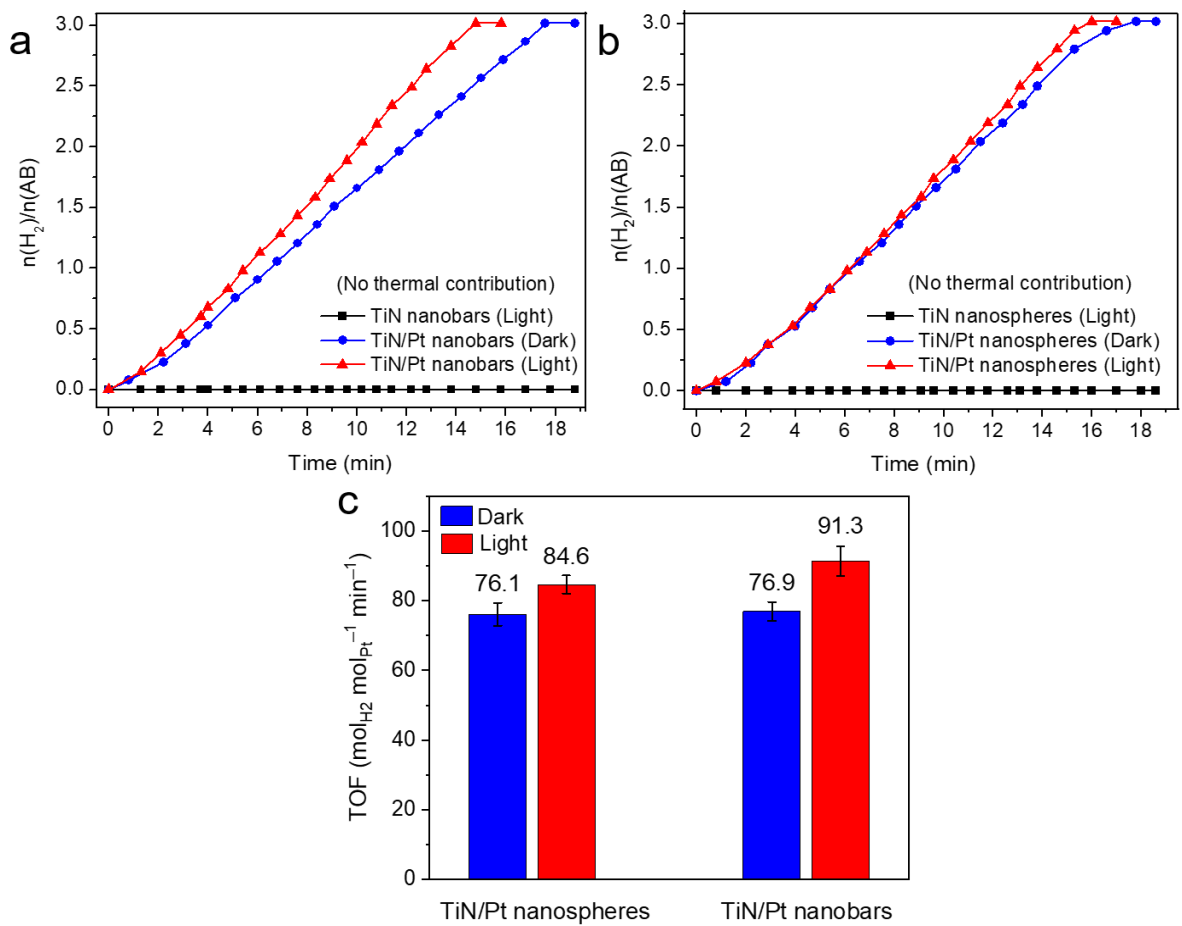

**Fig. S31 Comparison of hydrogen evolution rates for two different TiN morphologies.** Reaction kinetic study of $NH_3BH_3$ dehydrogenation reaction catalyzed by TiN/Pt nanohybrids with two different morphology, such as (a) nanobars and (b) nanospheres. The starting temperature and the final temperature of the reaction solution were same in all the cases. Therefore, the rate enhancement is due only to the plasmonic hot electrons, with no effect from thermal heating. $n(H_2)/n(AB)$ represents the molar ratio between the evolved $H_2$ and utilized ammonia borane (AB). The stoichiometric ratio was three equivalent of hydrogen gas formed from one equivalent of ammonia borane. (c) Comparison of TOF values for both dark and light conditions plotted for two different morphologies. TiN nanobars show higher plasmon enhanced $H_2$ evolution rates as compared to TiN nanospheres. All reactions were carried out in the dark at room temperature and under 940 nm illumination with power density of 24 mW $cm^{-2}$. The error bar indicates the variation over three separate measurements.

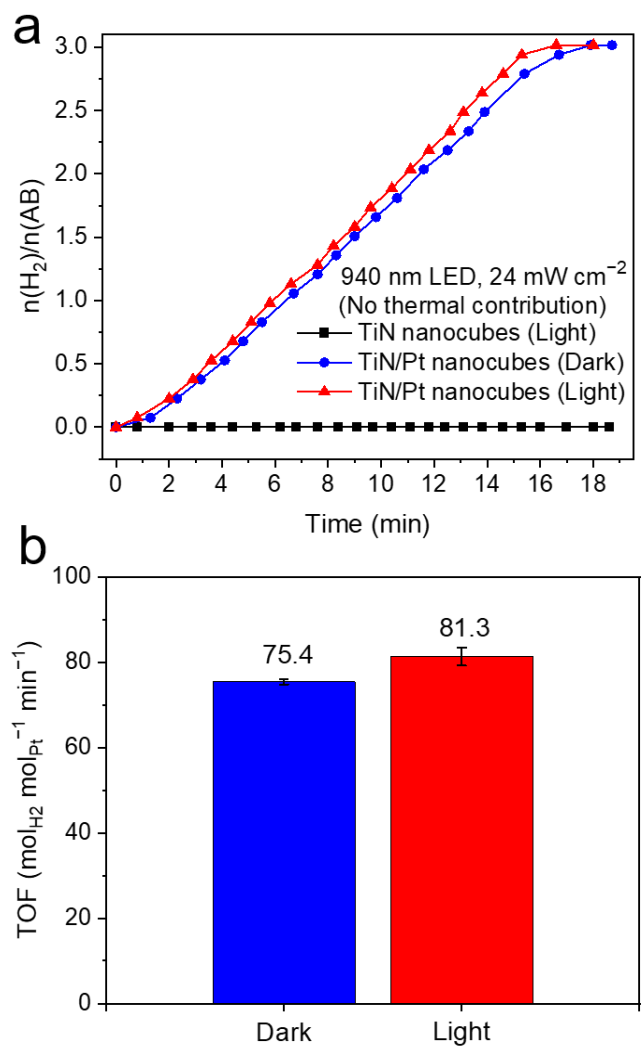

**Fig. S32 Photocatalytic performance of TiN/Pt nanocubes.** (a) Kinetics of $H_2$ evolution from ammonia borane with TiN/Pt nanocubes under dark at room temperature and under 940 nm illumination at a power density of 24 mW cm$^{-2}$. The starting temperature and the final temperature of the reaction solution were the same in the both cases. Therefore, the rate enhancement is due only to the plasmonic hot electrons, with no effect from thermal heating. (f) Comparison of TOF values for both dark and light conditions plotted for TiN/Pt nanocubes. The error bar indicates the variation over three separate measurements.

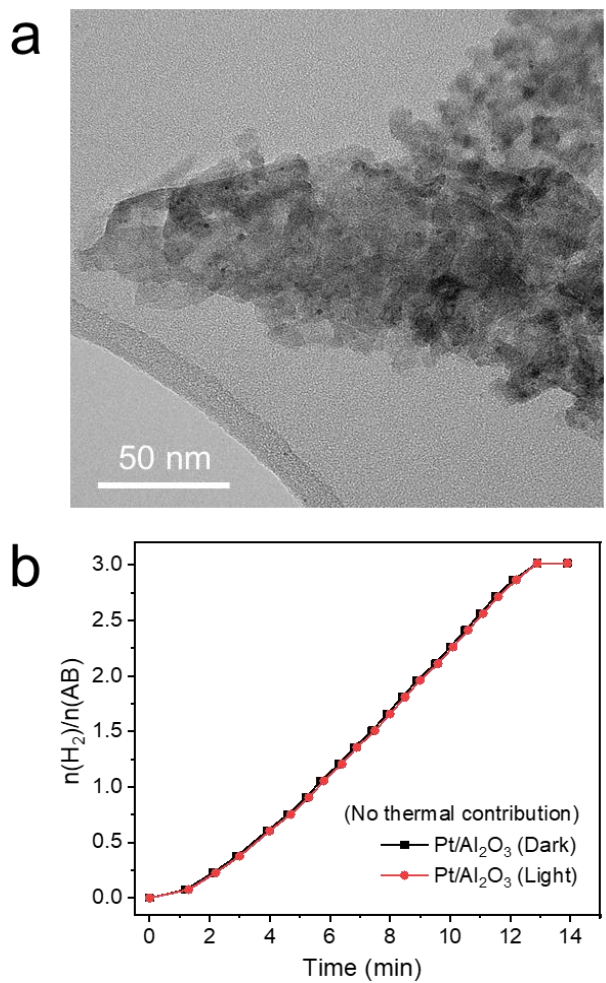

**Fig. S33 Photocatalytic performance of Pt/Al$_2$O$_3$ catalyst.** (a) TEM image of Pt/Al$_2$O$_3$ nanocatalyst. Highly disperse Pt nanocrystals (size: 2 ± 0.5 nm) can be seen on the Al$_2$O$_3$ matrix. (b) Kinetics of H$_2$ evolution from ammonia borane with Pt/Al$_2$O$_3$ under dark at room temperature and under 940 nm illumination at a power density of 24 mW cm$^{-2}$. The two-reaction kinetics are almost identical showing that Pt nanoparticles deposited on an inert support are not active under illumination.

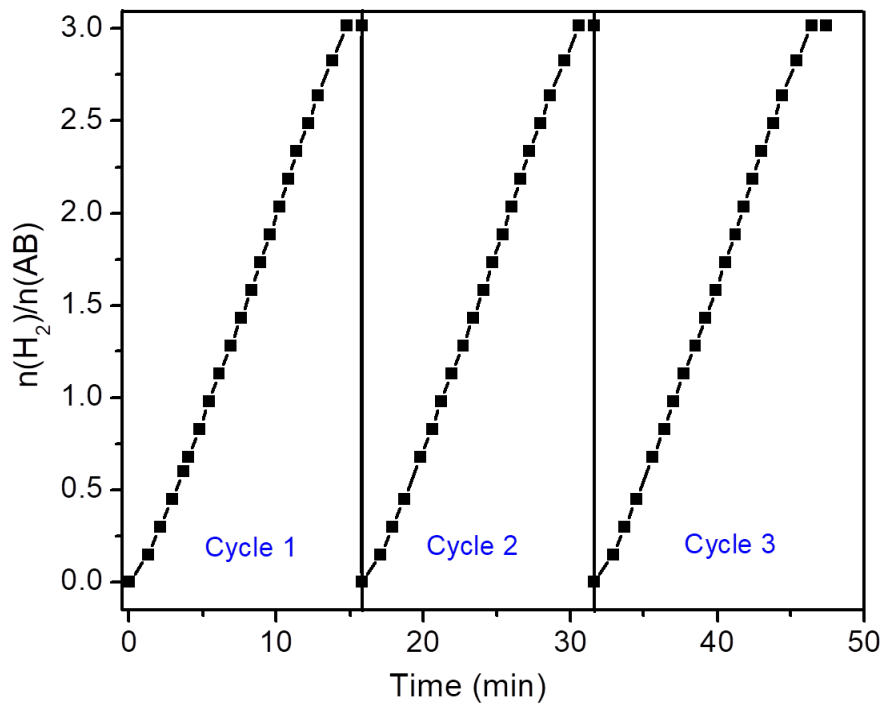

**Fig. S34 Recyclability test.** Recyclability test for the dehydrogenation of $NH_3BH_3$ over the TiN/Pt nanobars using 940 nm excitation at 24 mW $cm^{-2}$ light intensity.

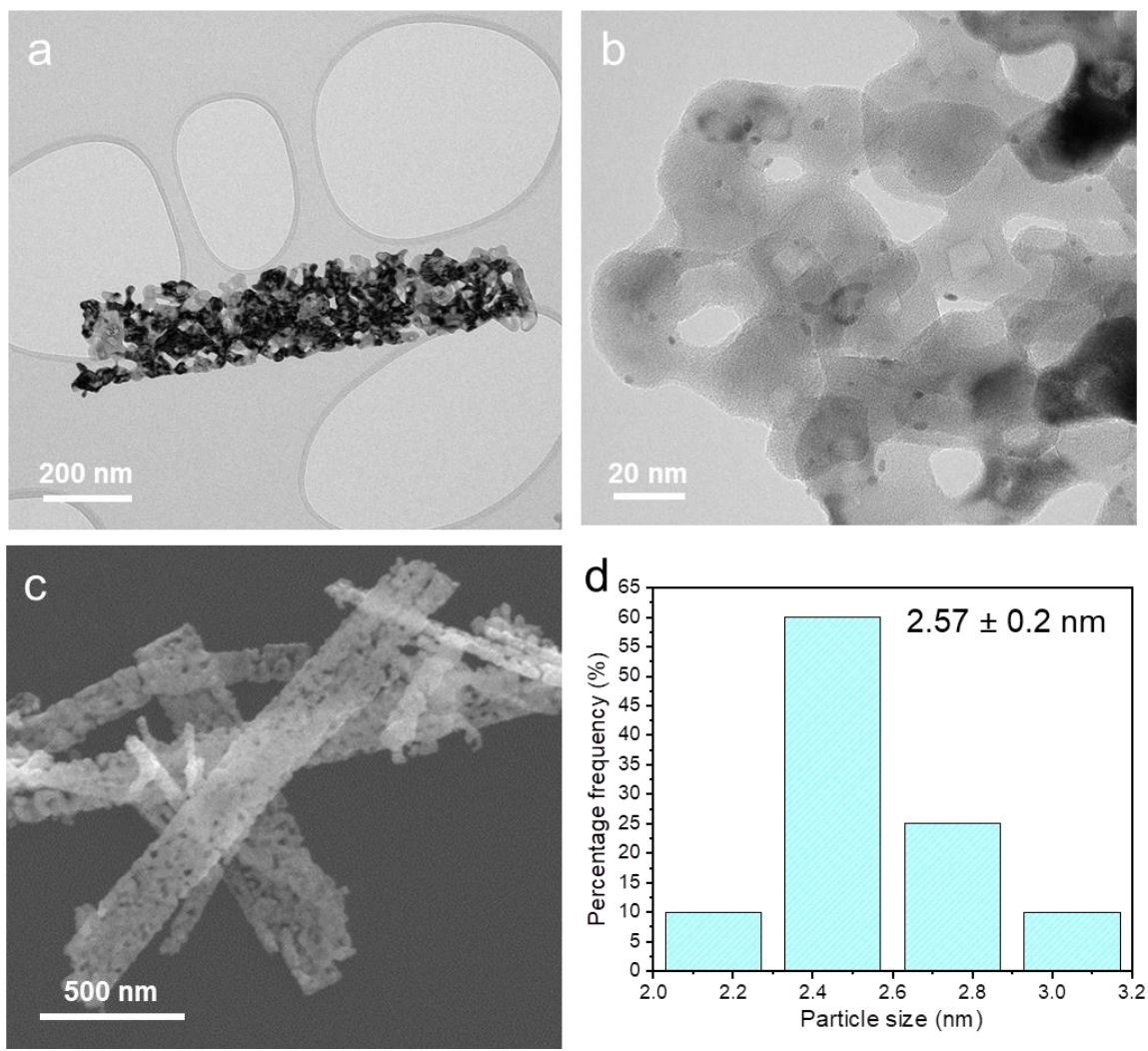

**Fig. S35 Stability test.** TEM images of (a) a single and (b) a magnified fragment of TiN/Pt nanobars after three catalytic cycles under 940 nm excitation at 24 mW cm$^{-2}$ light intensity. (c) SEM image of TiN/Pt nanobars after three catalytic cycles. (d) Size distribution histogram of the Pt nanocrystals obtained from 100 measurements after three catalytic cycles. The morphology of TiN nanobars and size distribution of Pt nanocrystals do not evidently change from the initial ones (i.e. before photocatalysis), highlighting the good stability of the catalyst.

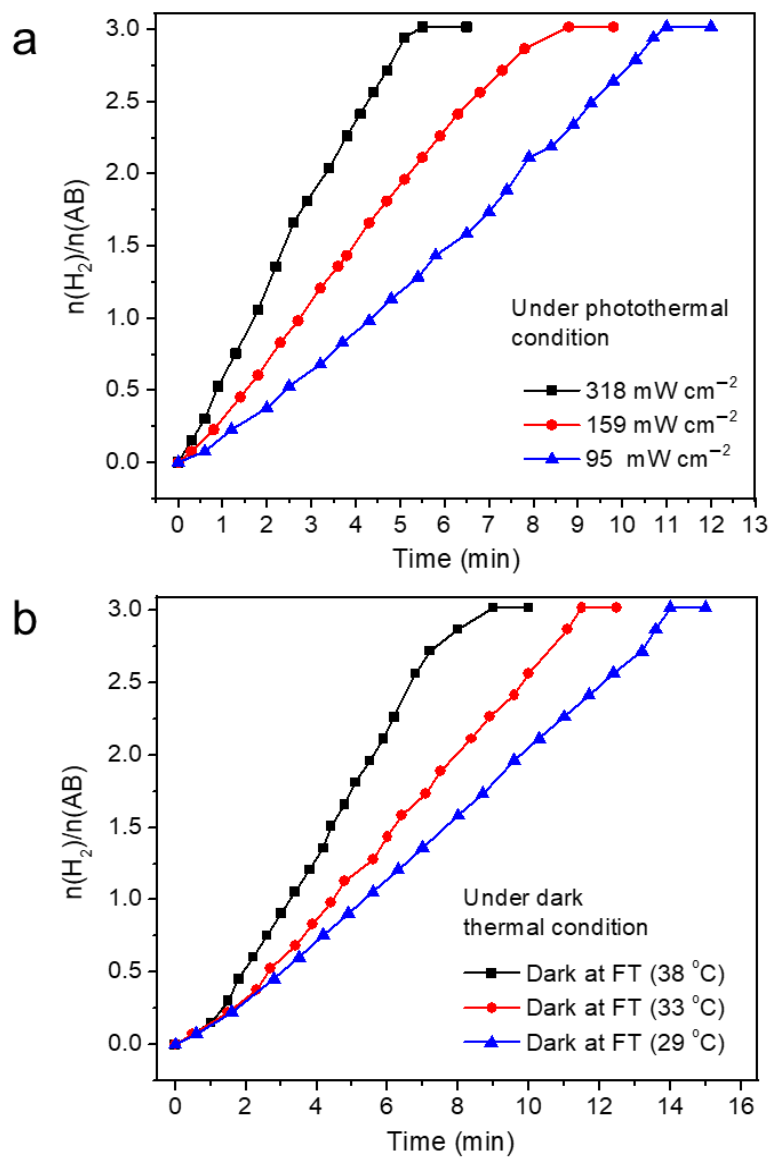

**Fig. S36 Hydrogen evolution kinetics from ammonia borane at different light intensity.** Kinetics of the NH$_3$BH$_3$ dehydrogenation reaction catalyzed by TiN/Pt nanobars under (a) different intensity at 940 nm excitation and (b) under different final temperatures (FTs) in dark condition.

**Table S4. Variation of TOF values under different conditions for TiN/Pt samples.**

| Condition | TOF ($mol_{H_2} mol_{Pt}^{-1} min^{-1}$) | Enhancement factor |
|---|---|---|
| TiN/Pt nanospheres (Dark) | 76.1 | 1.11 |
| TiN/Pt nanospheres (24 mW cm$^{-2}$, no thermal contribution) | 84.6 | |
| TiN/Pt nanobars (Dark) | 76.9 | 1.18 |
| TiN/Pt nanobars (24 mW cm$^{-2}$, no thermal contribution) | 91.3 | |
| TiN/Pt nanocubes (Dark) | 75.4 | 1.07 |
| TiN/Pt nanocubes (24 mW cm$^{-2}$, no thermal contribution) | 81.3 | |
| TiN/Pt nanobars (Dark thermal at 29 °C) | 95.7 | 1.28 |
| TiN/Pt nanobars (95 mW cm$^{-2}$, FT 29 °C) | 123.4 | |
| TiN/Pt nanobars (Dark thermal at 33 °C) | 117.4 | 1.30 |
| TiN/Pt nanobars (159 mW cm$^{-2}$, FT 33 °C) | 153.5 | |
| TiN/Pt nanobars | 151.0 | 1.60 |
| (Dark thermal at 38 °C) | | |
| TiN/Pt nanobars (318 mW cm$^{-2}$, FT 38 °C) | 242.3 | |

All the experiments were performed by starting at 22 °C. The TOF values were obtained by taking the average value from three separate measurements. The enhancement factor was computed as follow:

$$Enhancemnt\ factor = \frac{TOF_{Photothermal}}{TOF_{Dark\ thermal\ at\ FT}}$$

**Table S5. Literature study for catalytic activity comparison of different catalysts for H$_2$ evolution from ammonia borane both in the dark and light.** TOF values are expressed as $(mol_{H_2} mol_{Pt}^{-1} min^{-1})$. The total moles of Pt loaded on the catalyst were used for the TOF calculation in the comparison.

| Catalysts | T (°C) | TOF | Conditions | Reference |
|---|---|---|---|---|
| TiN/Pt nanobars | 22 | 76.9 | Dark | This work |
| TiN/Pt nanobars | 22 | 91.3 AQY: 120% | Under light (940 nm, 24 mW cm$^{-2}$) | This work |
| TiN/Pt nanobars | 38 | 242.3 $\Delta E_a = 0.28$ eV[#1] | Under light (940 nm, 318 mW cm$^{-2}$) | This Work |
| Ag nanorice /W$_{18}$O$_{49}$ nanowires | 15 | AQY: 4.03% | 1250 nm NIR light (Intensity not mentioned) | 19 |
| W$_{18}$O$_{49}$ nanowire/Carbon heterostructure | 15 | 0.14 μmol min$^{-1}$ | IR-light irradiation (λ > 750 nm) (Intensity not mentioned) | 20 |
| W$_{18}$O$_{49}$ nanowire /TiO$_2$ branched heterostructure | 15 | 0.62 μmol min$^{-1}$ | IR-light irradiation (λ > 750 nm) (Intensity not mentioned) | 21 |
| Ni$_x$Cu$_y$ nanoalloy/ carbon nitride nanosheets | 25 | 30.6 | λ > 420 nm 100 mW cm$^{-2}$ | S7 |
| Pt tipped Au nanorods | 25 | 981.3$^{\$}$ ΔE$_a$ = 0.10 eV$^{\#2}$ | λ > 320 nm 500 mW cm$^{-2}$ | S8 |
| Pt/CNT-O-HT | 25 | 50.6 | Dark | S9 |
| Pt/γ-Al$_2$O$_3$ | 30 | 24.7 | Dark | S10 |
| Pt/MIL-101 | 25 | 23.5 | Dark | S11 |
| Pt single atom on Ni/CNT | 25 | 33.3 | Dark | S12 |
| Pt@ZIF-8 | 20 | 16.6 | Dark | S13 |
| Pt/CNT | 25 | 23.6 | Dark | S14 |

\#1: Decrease of activation energy in the presence of 940 nm LED light with different intensity.

\#2: Decrease of activation energy in the presence of solar light (λ > 320 nm) with different intensity.

$: Pt loading not mentioned.

**Table S6. Activation energy ($E_a$) reduction comparison among a selection of different plasmon enhanced photocatalytic reactions recently reported in the literature.**

| Reaction | Plasmonic photocatalyst | $E_a$ reduction (eV) | Reference |
|---|---|---|---|
| Ammonia ($NH_3$) decomposition | Cu–Ru surface alloy nanocrystals supported on MgO | 0.94 | 7 |
| $H_2$ dissociation | Al octopods supported on γ-$Al_2O_3$ | 0.35 | S15 |
| Methanol steam reforming | Zn–Cu alloy nanocrystals supported on $SiO_2$ | 0.50 | S16 |
| Dry reforming of methane | Ni/$Ga_2O_3$ | 0.12 | S17 |
| Hydrodefluorination (HDF) of $CH_3F$ | Al−Pd nanocrystals supported on γ-$Al_2O_3$ | 0.21 | S18 |
| Fluorogenic oxidation reaction between nonfluorescent Amplex Red and $H_2O_2$ | Single Au nanorod | 0.11 | S19 |
| Dry reforming of methane | Pt−Au nanoalloy supported on $SiO_2$ | 0.73 | S20 |
| Carbon dioxide reduction | Rh/$TiO_2$ | 0.24 | S21 |

| Electrocatalytic hydrogen evolution reaction | Dumbbell-like Pt/Fe–Au nanorods | 0.19 | S22 |
| Methanol synthesis via $CO_2$ reduction | Cu/ZnO | 0.34 | S23 |

Kinetic Isotope Effect (KIE)

KIE experiments help to identify the rate determining step (RDS) of a reaction. Therefore, to understand KIE in the ammonia borane dehydrogenation reaction, we conducted the reaction in deuterium oxide ($D_2O$) instead of $H_2O$. Both of the reaction carried out under dark and using 940 nm excitation.

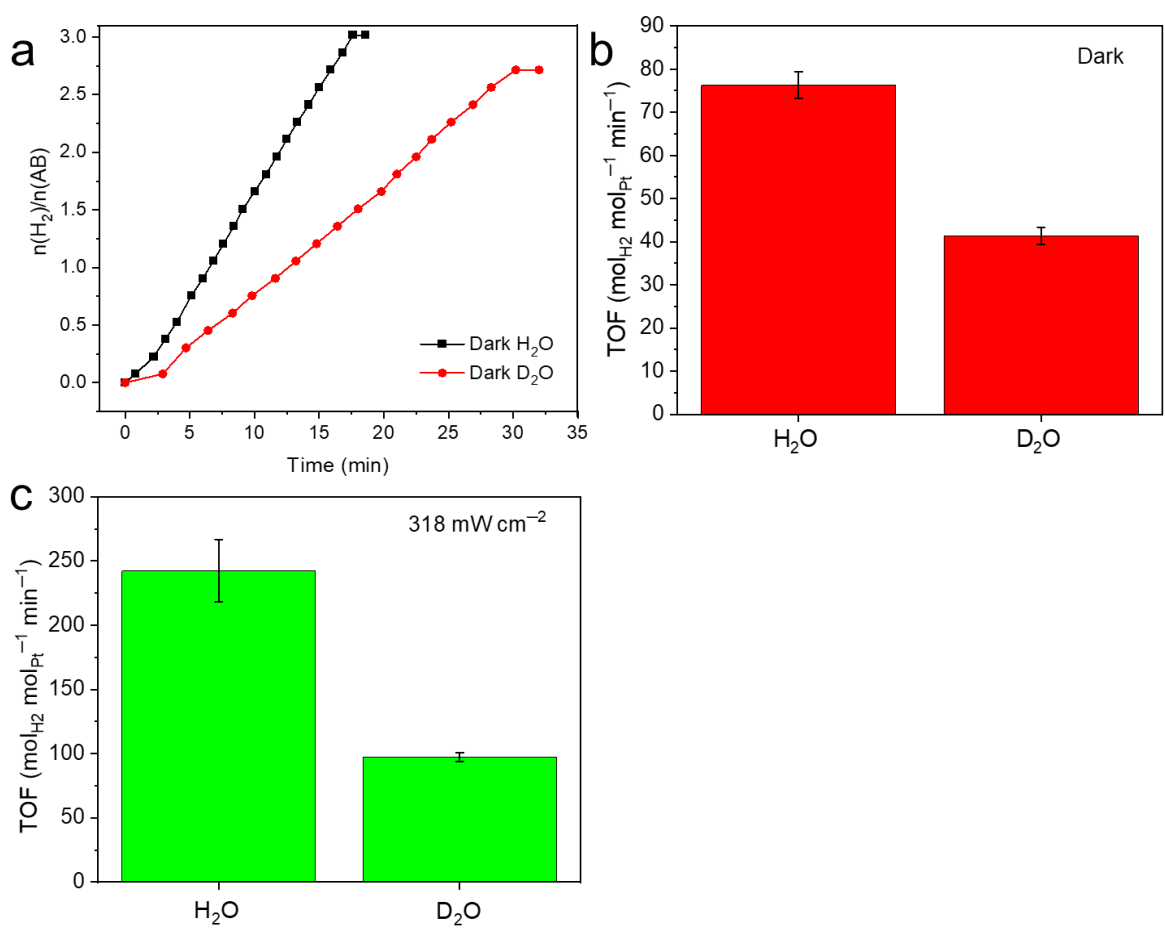

**Fig. S37 Kinetic Isotope Effect (KIE).** (a) Kinetics of $H_2$ evolution with TiN/Pt nanobars in $H_2O$ and $D_2O$ under dark conditions to study the KIE in the ammonia borane dehydrogenation reaction. Comparison of turnover frequency of $H_2$ evolution calculated under (b) dark at room temperature and (c) light (940 nm excitation at 318 mW cm$^{-2}$) conditions. The error bar indicates the variation over three separate measurements.

Considering TOF values from KIE experiments:

$$KIE_{dark} = \frac{TOF_{H_2O}(dark)}{TOF_{D_2O}(dark)} = \frac{76.3}{41.4} = 1.8$$

Under dark conditions, we obtained a KIE of 1.8, which strongly indicates that the cleavage of O–H bond in a water molecule is the rate-determining step during ammonia borane dehydrogenation reaction.

$$KIE_{light} = \frac{TOF_{H_2O}(light)}{TOF_{D_2O}(light)} = \frac{242.3}{97.5} = 2.4$$

A higher KIE value under light illumination (2.4) as compared to dark (1.8) confirms that a hot electron mechanism is playing a role in ammonia borane dehydrogenation using TiN/Pt nanobars.

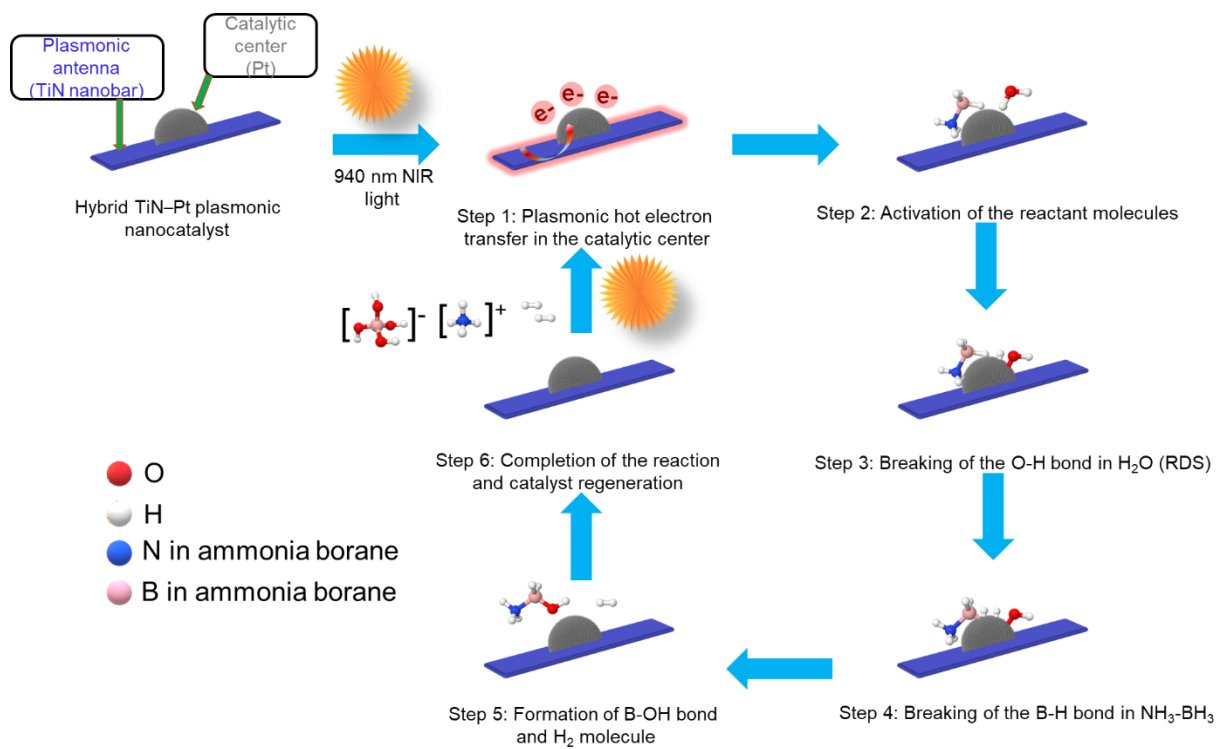

**Fig. S38 Hot electron driven mechanism.** Schematic representation of the hot electron driven $H_2$ evolution mechanism from $NH_3BH_3$ dehydrogenation over TiN/Pt nanobars under 940 nm NIR light irradiation.

**Plasmonic surface enhanced infrared absorption spectroscopy**

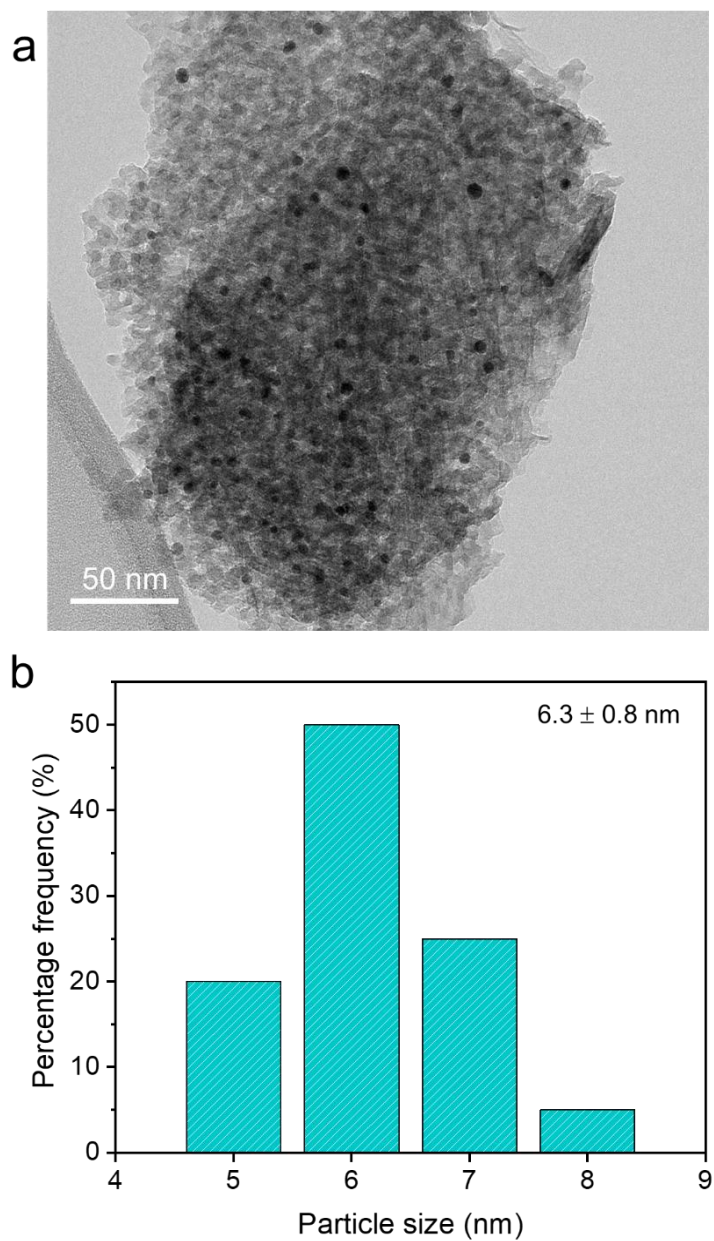

**Fig. S39 Morphological characterization of Au nanoparticles supported on Al$_2$O$_3$.** (a) TEM images of Au (6 wt.%)/Al$_2$O$_3$. (b) Size distribution histogram of Au nanocrystals deposited on Al$_2$O$_3$ support obtained from 100 measurements.

**Table S7. Surface enhanced infrared absorption spectroscopy.** SEIRA signals of furfural deposited on plasmonic TiN nanobars and $Al_2O_3$ at different wavenumbers (wavelengths) and the corresponding SEIRA enhancement factor obtained by dividing the signals obtained using the plasmonic TiN nanobars with the ones detected over an inert $Al_2O_3$ substrate.

| Wavenumber (cm$^{-1}$) | Wavelength (nm) | Absorption on TiN nanobars | Absorption on $Al_2O_3$ | SEIRA enhancement |
|---|---|---|---|---|
| 3334 | 2999 | low | n.d. | - |
| 3128 | 3197 | 1.56×10$^{-3}$ | n.d. | - |
| 2803 | 3568 | 1.46×10$^{-3}$ | n.d. | - |
| 2711 | 3689 | 9.85×10$^{-4}$ | 1.33×10$^{-5}$ | 74 |
| 1670 | 5988 | 1.99×10$^{-2}$ | 7.79×10$^{-4}$ | 25 |
| 1566 | 6386 | 5.45×10$^{-3}$ | 5.55×10$^{-5}$ | 98 |
| 1562 | 6402 | 5.91×10$^{-3}$ | 2.49×10$^{-5}$ | 238 |
| 1388 | 7205 | 4.79×10$^{-3}$ | 2.43×10$^{-4}$ | 20 |
| 1365 | 7326 | 2.83×10$^{-3}$ | 6.25×10$^{-5}$ | 45 |
| 1273 | 7855 | 3.05×10$^{-3}$ | 2.03×10$^{-4}$ | 15 |
| 1242 | 8052 | 1.92×10$^{-3}$ | 6.14×10$^{-5}$ | 31 |
| 1223 | 8177 | 1.67×10$^{-3}$ | 5.89×10$^{-5}$ | 28 |
| 1149 | 8703 | 3.03×10$^{-3}$ | 1.22×10$^{-4}$ | 24.73 |
| 1076 | 9294 | 2.01×10$^{-3}$ | n.d. | - |
| 1014 | 9862 | 6.11×10$^{-3}$ | 2.29×10$^{-4}$ | 26.66 |
| 926 | 10799 | 2.93×10$^{-3}$ | 8.79×10$^{-5}$ | 33.38 |
| 879 | 11377 | 3.06×10$^{-3}$ | 4.10×10$^{-4}$ | 7.47 |
| 833 | 12005 | 2.31×10$^{-3}$ | 2.62×10$^{-4}$ | 8.82 |
| 748 | 13369 | 1.42×10$^{-2}$ | 4.09×10$^{-4}$ | 34.70 |